\RequirePackage{fix-cm}
\documentclass[twocolumn]{svjour3}      
\smartqed  
\usepackage{newtxtext, newtxmath}      
\usepackage{latexsym}
\usepackage{booktabs}
\usepackage{cite}
\usepackage{url}
\usepackage{balance}
\usepackage{flushend}
\usepackage[misc,geometry]{ifsym}

\usepackage{lipsum}
\usepackage{rotating}
\usepackage{pdflscape}      
\usepackage{longtable}       
\usepackage{afterpage}
\usepackage{tabularx}
\usepackage{adjustbox}
\usepackage{afterpage}
\usepackage{wrapfig}
\usepackage{multirow}
\usepackage{placeins}
\usepackage{float}

\usepackage{graphicx}

\usepackage{xcolor}

\usepackage[symbol]{footmisc}

\usepackage{booktabs,dcolumn}                     
\usepackage[figureposition=bottom]{caption}                              
 \makeatletter
  \providecommand*\setfloatlocations[2]{\@namedef{fps@#1}{#2}}
\makeatother

\makeatletter
  \@addtoreset{table}{section}
\makeatother
\renewcommand*\thetable{\arabic{section}.\arabic{table}}

\newcolumntype{d}[1]{D{.}{.}{#1}}

\usepackage{makecell}

\renewcommand{\thetable}{\arabic{table}}
\usepackage{threeparttable}

\usepackage{booktabs}
\usepackage{tabularx}
\usepackage{adjustbox}
\usepackage{rotating}
\usepackage{caption}
\usepackage{geometry}
\usepackage{placeins}

\usepackage[utf8]{inputenc}
\usepackage[T1]{fontenc}

 \journalname{International Journal of Social Robotics}
\begin{document}

\title{Social Robots for People Living with Dementia: A Scoping Review on Deception from Design to Perception}

\titlerunning{Social Robots for People Living with Dementia: A Scoping Review on Deception from Design to Perception}        
\author{Fan Wang \and Giulia Perugia \and Yuan Feng \and Wijnand IJsselsteijn}

\authorrunning{F. Wang et al.} 

\institute{Fan Wang \at
              Department of Industrial Engineering and Innovation Science, Eindhoven University of Technology, the Netherlands \\
              \email{f.wang3@tue.nl}
           \and
           Giulia Perugia \Letter  \at
              Department of Industrial Engineering and Innovation Science, Eindhoven University of Technology, the Netherlands \\
              \email{g.perugia@tue.nl}
           \and
           Yuan Feng \at
              Department of Industrial Design, Northwestern Polytechnical University, China \\
              \email{y.feng@nwpu.edu.cn}
           \and
           Wijnand IJsselsteijn \at
              Department of Industrial Engineering and Innovation Science, Eindhoven University of Technology, the Netherlands \\
              \email{w.a.ijsselsteijn@tue.nl}
}

\date{Received: date / Accepted: date}

\maketitle

\begin{abstract}
As social robots are increasingly introduced into dementia care, their embodied and interactive design may blur the boundary between artificial and lifelike entities, raising ethical concerns about robotic deception. However, it remains unclear which specific design cues of social robots might lead to social robotic deception (SRD) in people living with dementia (PLwD), and which perceptions and responses of PLwD might indicate that SRD is taking place. To address these questions, we conducted a scoping review of 26 empirical studies reporting PLwD interacting with social robots. We identified three key design cue categories that might contribute to SRD and one that might break the illusion. However, the available literature does not provide sufficient evidence to determine which specific design cues lead to SRD. Thematic analysis of user responses reveals six recurring patterns in how PLwD perceive and respond to social robots. However, conceptual limitations in existing definitions of robotic deception make it difficult to identify when and to what extent deception actually occurs. Building on the results, we propose a dual-process interpretation that clarifies the cognitive basis of false beliefs in human–robot interaction and distinguishes SRD from anthropomorphism or emotional engagement.

\end{abstract}

\section{Introduction}\label{sec:intro}

More than 55 million people worldwide have a diagnosis of dementia, and the number is expected to reach 78 million by 2030 due to the aging population \cite{alzheimers_stats}. Dementia is a term that encompasses several diseases (e.g., Alzheimer's disease, vascular dementia, frontotemporal dementia) that over time damage nerve cells and the brain, typically leading to deterioration in cognitive function and changes in behavior, emotion, and motivation \cite{who_dementia}. 

Social robots have emerged as a promising technological intervention to enhance the quality of life and psychological well-being of people living with dementia (PLwD) \cite{abdollahiPilotStudyUsing2017a}. Research has demonstrated their potential in mitigating symptoms such as agitation, anxiety, and depression \cite{joransonChangeQualityLife2016a, khoslaAssistiveRobotEnabled2014a, laneEffectivenessSocialRobot2016a, shibataTherapeuticSealRobot2012}, as well as promoting social engagement \cite{chanPromotingEngagementCognitively2010, mannionIntroducingSocialRobot2020a, perugiaElectrodermalActivityExplorations2017a, wadaRobotTherapyCare2007, feng2022context}.
Beyond questions of effectiveness and acceptance, however, social robots raise a range of ethical concerns
\cite{o2019robots}.
Among these concerns, deception is one particularly contentious and complex to unpack \cite{arkin2011moral, broadbentInteractionsRobotsTruths2017, hancockMetaAnalysisFactorsAffecting2011, olesonAntecedentsTrustHumanrobot2011, saetraSocialRobotDeception2021}. 

Social robotic deception (SRD), which Danaher defines as the phenomenon whereby ``a robot, as an artificial agent, creates a misleading impression through its representations or signals'' \cite{danaherRobotBetrayalGuide2020}, may lead users to overestimate robots' capabilities and emotional states, creating illusions of lifelikeness, sentience, and cognition, especially in cognitively vulnerable user groups such as people living with dementia (PLwD) \cite{sharkeyWeNeedTalk2021}. At the same time, it may be precisely the overestimations and illusions brought about by SRD that enable social robots to achieve the therapeutic effects highlighted in literature.
Although some researchers have briefly mentioned issues such as misperception, attachment, and infantilization when social robots are used with PLwD \cite{kohImpactsLowcostRobotic2021, lengPetRobotIntervention2019, palmier2024identification}, there remains a significant gap in understanding how SRD arises and unfolds in this population. Existing literature reviews have mainly focused on the feasibility, effectiveness, and acceptance of social robots in dementia care \cite{ghafurianSocialRobotsCare2021, hirtSocialRobotInterventions2021, hungBenefitsBarriersUsing2019, kohImpactsLowcostRobotic2021, lengPetRobotIntervention2019, luEffectivenessCompanionRobot2021, ongEffectivenessRobotTherapy2021, satheriiiUseHumanoidRobot2021, whelanFactorsAffectingAcceptability2018}, devoting little attention to which design cues are embedded in social robots for PLwD, and how PLwD perceive, interpret, and make sense of social robots.

This scoping review is situated within the delicate balance between social robots' ethical shortcomings and their therapeutic benefits, and aims to add nuance to the issue of SRD by describing what design and interactive robot cues have been embedded in social robots for PLwD by the Human-Robot Interaction (HRI) scholarship, and which perceptual responses they have generated in PLwD. While our results are descriptive in nature, as per tradition of scoping reviews\cite{munn2018systematic}, in our discussion we take on a more critical stance and pinpoint which of the many identified cues might contribute to SRD in PLwD, and which perceptions in PLwD might indicate that SRD is actually taking place.

\subsection{Understanding Deception in HRI from Design and User Perspectives}\label{subsec1.2}
Many studies in HRI draw on the Computers Are Social Actors (CASA) paradigm, which suggests that humans respond to robots as they would to other social agents \cite{nassComputersAreSocial1994, gambinoBuildingStrongerCASA2020, reeves1996media}.
Within this context, ethical concerns about robotic deception have sparked debates on its identification and characteristics. 
From a design perspective, several scholars define robotic deception by emphasizing the significance of robot design features.
Wallach and Allen \cite{allen2009moral} suggest that enabling robots to respond with human-like social cues is a form of deception.
Similarly, Matthias \cite{matthias2015robot} contends that it is deception when a robot suggests a mental or emotional capability that it does not possess.
Sharkey and Sharkey \cite{sharkey2011children} further assert that creating the illusion of mental states in robots is inherently deceptive, as robots lack thoughts or subjective experiences.
Following the design perspective, this scoping review attempts to unpack those design cues embedded into social robots for PLwD which could disguise the artificial nature of the robot. As such, we posit the following research question, \textbf{RQ1}:
\begin{itemize}
    \item[\textbf{a.}] What design cues are embedded in social robots for people living with dementia? 
    \item[\textbf{b.}] Which of these cues might lead to social robotic deception in people living with dementia?
\end{itemize}

Furthermore, growing discussions highlight the need to identify robotic deception through user perception and response.
Coeckelbergh \cite{coeckelbergh2018describe} emphasizes the role of human subjectivity in interpreting robotic behaviors based on their relations and experience with robots, arguing that deception is relational and co-created by user, robot, and broader context.
Sharkey et al. \cite{sharkeyRobotsHumanDignity2014, sharkeyGrannyRobotsEthical2012, sharkeyWeNeedTalk2021} highlight the importance of focusing on the deceived rather than the deceiver, discussing how false beliefs can arise even without intentional deception. 
Grodzinskii et al. \cite{grodzinsky2015developing} argue that robotic deception occurs when people interpret a robot's behavior as indicative of a human or other biological life form.
Sorell and Draper \cite{sorell2017second} narrow the definition, stating that only when people are explicitly misled by the design of a robot and believe it to be a genuine human or animal, it is robotic deception.
Hence understanding how social robots are experienced and interpreted by PLwD can provide evidence of whether robotic deception is occurring or not.
Following this line of thought, this literature review analyzes PLwD's perception of social robots to pinpoint potential misleading impressions arising from robots' behavior and appearance. Hence, we formulate a second research question, \textbf{RQ2}:
\begin{itemize}
    \item[\textbf{a.}] What perceptions and responses do social robots elicit in people living with dementia? 
    \item[\textbf{b.}] Which of these perceptions might indicate that social robotic deception is actually taking place?
\end{itemize}

Existing definitions of robotic deception are largely grounded in philosophical argumentation. As a result, they often remain disconnected from the empirical literature on how users actually perceive and respond to robots in practice. This gap is particularly consequential in dementia care, where cognitive vulnerability and relational care contexts may shape how robotic behaviors are interpreted.
To bridge this gap, Esposito et al. \cite{esposito2024deception} suggest that future research should integrate psychological theories and methods into the study of robotic deception, ensuring a more structured and empirical approach. 
With this scoping review, we aim to extend this trajectory through the critical analysis of empirical evidence in dementia care. 
We primarily adopt the definition of SRD proposed by Danaher \cite{danaherRobotBetrayalGuide2020} that links the design of robots with perception of users. We also draw on insights from Coeckelbergh \cite{coeckelbergh2018describe}, who emphasizes the relational and contextual nature of human–robot interaction.
Grounding our analysis in these perspectives, we attempt to identify which design and interactive robot cues might lead to SRD in PLwD (\textbf{RQ1}) and which perceptions might reveal that SRD is actually taking place (\textbf{RQ2}). It is important to note that since the current analysis is dependent on the details disclosed by the authors of the reviewed papers, it will not always be possible to map design and interactive robot cues into PLwD's perceptions and responses.

\subsection {Understanding Deception Within Dementia Care}
Deception is a central ethical issue, not only in Social Robotics but also in Dementia Care \cite{dayPeopleDementiaFind2011, huangDeceptionDefensibleDementia2022, schermerNothingTruthTruth2007}, as caregivers often face the difficult choice of correcting misconceptions, lying, or distracting PLwD \cite{caseyTellingGoodWhite2020, turner2017use}. In this context, deception is generally understood as intentional, aligned with definitions such as ``all that we do or do not do, say or do not say, with the intention of misleading others'' \cite{schermerNothingTruthTruth2007}.

Prior research has distinguished between different forms and degrees of deception in dementia care.
For instance, Dresser \cite{dresser2021tangled} proposes a hierarchy
ranging from distraction and redirection -- which focus on acknowledging PLwD's false beliefs and other minor distortions that fit into their subjective realities -- to deceptive claims and outright lies -- which intentionally contradict truth, and are only justified as a last resort to protect PLwD from physical and psychological harm. 
For instance, a caregiver may go along with a person's mistaken belief that they are their office assistant, rather than correcting them, to avoid disorientation and distress. In another case, a man whose wife became upset every time he said goodbye chose instead to tell her he was going shopping. This allowed her to feel reassured that he would return soon, reducing her anxiety.

Dementia-specific prosocial lies, which are intended to mislead but benefit PLwD, can sometimes be considered more ethical than honesty \cite{levine2014liars, levine2022community}. For example, caregivers might falsely inform PLwD that their family is coming soon to encourage them to get dressed or clean up. 
These discussions primarily frame deception from a practical perspective, in which ethical judgments are mainly made on a case-by-case basis by human caregivers \cite{wang2026following}.

Another major strand of literature addresses the acceptability of deception in dementia care. 
 Researchers conclude that morally defensible deception should prioritize dignity, align with cognitive capacities \cite{gastmans2013dignity, huangDeceptionDefensibleDementia2022, dresser2021tangled}, and consider the possibility of recovery \cite{abdool2015ethics}.
However, these assessments are most often grounded in the perspectives of caregivers and researchers, with limited direct engagement with how PLwD themselves experience or interpret deceptive practices.
Although some studies have attempted to capture PLwD's views \cite{caseyTellingGoodWhite2020, dayPeopleDementiaFind2011}, they primarily rely on post-hoc reflections. 
As a result, decisions on deception in dementia care are commonly made on behalf of PLwD rather than emerging from a negotiation with them.

Existing literature provides a lens to consider the morality and ethicality of certain HRI practices in dementia care. Tummers-Heemels et al.\cite{tummers2021between} offer a nuanced ethical reflection on the continuum between benevolent lies and harmful deception in the broader context of dementia care technologies. They emphasize that deception in care is not inherently unethical but must be evaluated in light of its intention, transparency, and impact on the person's dignity and autonomy. Their work highlights the importance of situational and relational ethics, suggesting that the acceptability of deception depends not only on outcomes, but also on the degree to which PLwD are meaningfully included in shaping their own care context. 
This is particularly relevant for social robots, whose embodied presence and interactive behaviors may systematically shape users' beliefs and expectations over time.

Taken together, prior work on deception in dementia care provides important ethical foundations but remains largely focused on human caregivers and practical decision-making. 
In the context of social robots, discussion around deception requires a shift in focus: from the intentions of caregivers or designers to the preferences of PLwD themselves, in line with \cite{ijsselsteijnWarmTechnologyNovel2020, suijkerbuijk2019active}. 
This scoping review provides a structured synthesis that could help with this endeavor. It describes which design cues are embedded in social robots for PLwD and when they can lead to deception. Moreover, it identifies the perceptions and responses reported among PLwD during interaction, and which of them are representative of SRD. By bringing these strands of literature together, it clarifies how robot design and user perceptions can be interpreted in discussions of SRD in dementia care.


\section{Methodology}\label{sec:method}
In order to identify the papers to include in this scoping review, we performed an electronic search in the following six databases: Web of Science, IEEE Xplore, Science Direct, Scopus, ACM Digital Library, and PubMed. We limited our search to the past ten years, as the use of social robots for PLwD has increased significantly during this period. Only papers written in English were included. We used the following search strings:
\begin{itemize}
    \item Search string1: ``Robot*'' AND (``Cognitive* Impairment'' OR Dementia* OR Alzheimer*)
    \item Search string2: ``Robot'' AND (``Cognitive Impairment'' OR Dementia OR Alzheimer)
\end{itemize}

The search yielded a list of 991 papers of which:
\begin{itemize}
    \item 365 from Web of Science (search string 1)
    \item 170 from Scopus (search string 1)
    \item 115 from IEEE Xplore (search string 1)
    \item 102 from PubMed (search string 1)
    \item 96 from ACM Digital Library (search string 1)
    \item 143 from Science Direct (search string 2)
\end{itemize}

\begin{figure*}[h]%
\centering
\includegraphics[width=\linewidth]{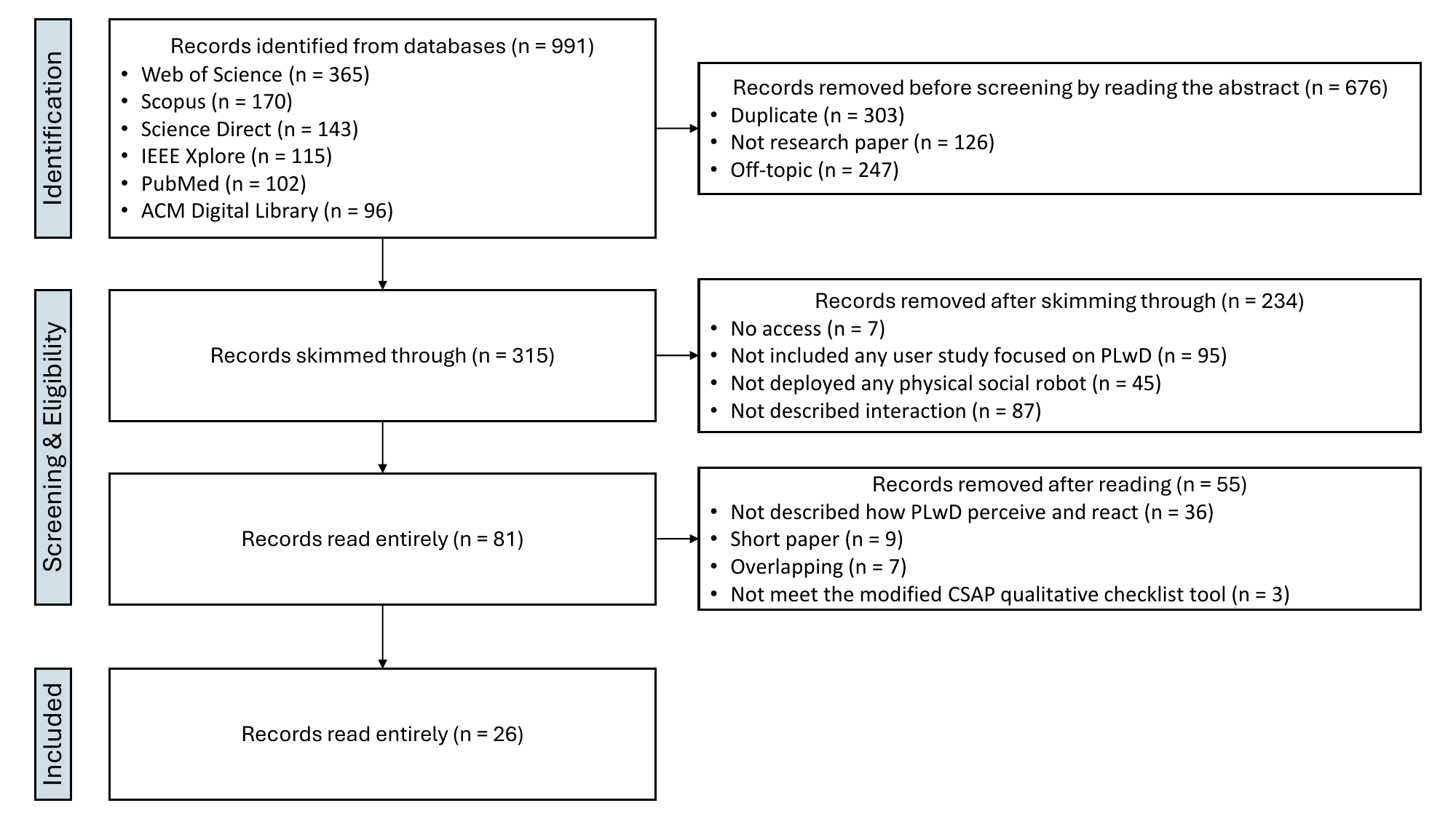}
\caption{PRISMA diagram for the paper selection pipeline}\label{PRISMA diagram}
\end{figure*}

\begin{figure*}[h]%
\centering
\includegraphics[width=\linewidth]{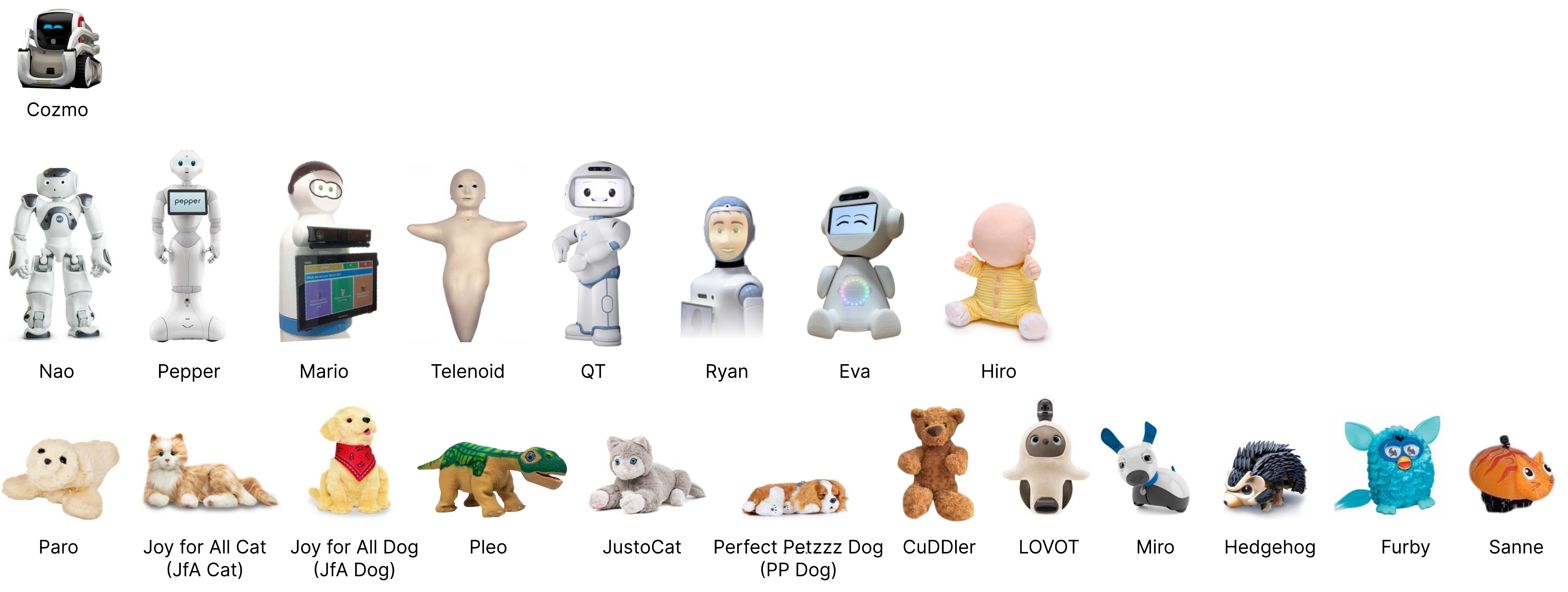}
\caption{The robots utilized in the included studies. The first row shows abstract robot, the second row shows anthropomorphic robots, and the third row shows biomorphic robots. The robots are presented in descending order based on their frequency of use, from left to right.}\label{Robots}
\end{figure*}

\begin{table*}[h]
\vspace*{\fill} 
\centering
\caption{Overview of the structures and HRI details in the included studies (NS = not specified). `-*'' in the Length of the HRI column represents one-session activity; `-*'' in the Duration of the sessions column represents daily-based activities.}
\label{structures}
\begin{tabularx}{\textwidth}{>{\arraybackslash}p{3.5cm} >{\arraybackslash}p{1.8cm} >{\arraybackslash}p{4.5cm} >{\arraybackslash}X}
\toprule
\textbf{Author} & \textbf{Structure} & \textbf{Length of the HRI} & \textbf{Duration of the sessions} \\
\midrule
Abdollahi et al. (2022) \cite{abdollahi2022artificial} & structured & 3 weeks, twice a week & about 15 minutes \\
Bradwell et al. (2022) \cite{bradwellImplementingAffordableSocially2022} & free interaction & 4 months; 8 months & 15-minute sessions to 24/7 "adoption" \\
Bradwell et al. (2021) \cite{bradwellUsercenteredDesignCompanion2021} & free interaction & - & about 30 to 45 minutes in total \\
Casey et al. (2020) \cite{casey2020perceptions} & free interaction & 2 months & average of 41.3 minutes across sites \\
Cruz-Sandoval et al. (2020) \cite{cruz2020social} & structured & 7 weeks & about 30 minutes \\
Dinesen et al. (2022) \cite{dinesenUseSocialRobot2022} & free interaction & IS: over 4 weeks, twice a week; \newline GS: over 12 weeks, twice a week & IS: about 20–30 minutes; \newline GS: 30–45 minutes \\
Feng et al. (2019) \cite{feng2019livenature} & free interaction & 4 weeks, once per week & up to 20 minutes \\
Gustafsson et al. (2015) \cite{gustafsson2015using} & free interaction & 7 weeks & - \\
Hsu et al. (2023) \cite{hsu2023co} & structured & 5 weeks, once per week & 15 to 45 minutes \\
Hung et al. (2021) \cite{hungExploringPerceptionsPeople2021} & free interaction & 2 weeks--6 months or longer, 2--4 times & about 20 to 30 minutes \\
Inoue et al. (2021) \cite{inoue2021exploring} & free interaction & 1 to 3 months, over 3 times per week & 15 to 180 minutes \\
Joshi et al. (2019) \cite{joshi2019robots} & free interaction & over 4 months, 4 sessions & NS \\
Kuwamura et al. (2016) \cite{kuwamuraCanWeTalk2016} & free interaction & 3 months, once or twice a week & up to 15 minutes \\
Marchetti et al. (2022) \cite{marchetti2022pet} & free interaction & 3 days & 7 to 20 minutes, 2:07 hours in total \\
Moyle et al. (2019) \cite{moyle2019using} & free interaction & 10 weeks, 3 times per week & 15 minutes \\
Moyle et al. (2016) \cite{moyleWhatEffectDoes2016} & free interaction & 5 weeks, 3 times per week & 30 minutes \\
Pike et al. (2021) \cite{pikeRobotCompanionCats2021} & free interaction & 3 months & - \\
Pu et al. (2020) \cite{puHowPeopleDementia2020} & free interaction & 6 weeks, 5 days per week & 30 minutes \\
Raß et al. (2023) \cite{rassInvestigatingPotentialImpacts2023} & structured & 2 months, 8 times per week & NS \\
Rouaix et al. (2017) \cite{rouaixAffectiveEngagementIssues2017} & structured & - & 35 sessions in total, average of 22.15 minutes \\
Sarabia et al. (2018) \cite{sarabia2018assistive} & free interaction & 5 days & 1st interaction: average of 7 min 10 s; \newline 2nd interaction: average of 3 min 23 s \\
Sumioka et al. (2021) \cite{sumioka2021minimal} & free interaction & - & up to 5 minutes \\
Tanioka et al. (2021) \cite{tanioka2021development} & free interaction & - & about 20 to 30 minutes \\
Thunberg et al. (2020) \cite{thunbergRobotPetsDecrease2020} & free interaction & RC: 9 months; \newline RD: 3 months & NS \\
Whelan et al. (2020) \cite{whelanEffectsMARIOSocial2020} & free interaction & 5 weeks, 3 times per week & NS \\
Yamazaki et al. (2014) \cite{yamazakiAcceptabilityTeleoperatedAndroid2014} & free interaction & - & about 2 hours \\
\bottomrule
\end{tabularx}
\vspace*{\fill} 
\end{table*}

\begin{table*}[h]
\vspace*{\fill} 
\centering
\caption{Overview of the structured activities in the included studies.}
\label{activities-only}
\begin{tabularx}{\textwidth}{>{\arraybackslash}p{3.5cm} >{\arraybackslash}X}
\toprule
\textbf{Author} & \textbf{Activities} \\
\midrule
Abdollahi et al. (2022) \cite{abdollahi2022artificial} & Conversational interaction with robots on preset topics, such as family, music, movies, etc. \\
Cruz-Sandoval et al. (2020) \cite{cruz2020social} & Robot-assisted therapeutic session including elements of musicotherapy, reminiscence, cognitive games (complete to wisdom sayings), and relaxation. \\
Hsu et al. (2023) \cite{hsu2023co} & Co-design workshops including discovering through storytelling, building a robot through hearing, dressing up the robot, dancing, and refective storytelling. \\
Raß et al. (2023) \cite{rassInvestigatingPotentialImpacts2023} & Robot-assisted group sessions for physical activities and conversations over biographics of the residents between the residents and the professional caregiver. \\
Rouaix et al. (2017) \cite{rouaixAffectiveEngagementIssues2017} & Robot-assisted psychomotor therapy, including guided motor exercises, cognitive stimulation, body expression activities, and personalized interactions. \\
\bottomrule
\end{tabularx}
\vspace*{\fill} 
\end{table*}

In the first exclusion round, the articles obtained from the search were imported into Rayyan, an online review platform. Six hundred and eighty-eight papers were left after 303 duplicate papers were removed based on their titles and DOIs. The first author read the abstracts of all 688 papers and excluded 373 of them. These excluded papers were review articles or articles other than research articles, such as prefaces, keynote abstracts or posters (N = 126) or articles that did not fit the topic of our review (N = 247), for example, articles (1) employing technologies other than social robots or therapies without social robots (N = 100) or (2) focusing on end users other than PLwD (N = 147). This screening process left us with 315 articles that were then independently screened by the first author and an independent researcher.

In the second round of exclusion, all articles were skimmed through and 234 of them were excluded because: (1) it was not possible to access the full article (N = 7); (2) the article did not present a qualitative or quantitative user study that focused on PLwD (N = 95); (3) the article did not focus on a physical social robot (N = 45); and (4) the article did not describe the interaction session(s) between a physical social robot and PLwD (N = 87).\footnote{By physical social robots, we refer to all embodied robotic systems that are physically present in the shared time and space with humans and are designed to engage in social interaction \cite{fong2003survey}.}
These exclusion criteria were applied because physical social robots uniquely suit PLwD by combining physical presence and affective communication modalities (e.g. facial expressions, gestures, body movements) with natural language interfaces, and our focus was on analyzing how PLwD perceive and respond to them.

After this step, we were left with 81 papers that were read by the two researchers independently in their entirety. The two researchers independently screened the same batch of articles and discussed their results during weekly meetings to ensure consistency.
During this round of screening, we excluded 52 articles that (1) did not describe how PLwD react and perceive social robots (N = 36); (2) were short articles, lacking sufficient methodological detail or empirical evidence (N = 9); or (3) were alternative versions of other articles already featured in the review (e.g. conference papers or previous drafts) (N = 7). 

A quality evaluation was conducted for the remaining articles using a modified Critical Appraisal Skills Programme (CASP) qualitative checklist tool \cite{longOptimisingValueCritical2020}. 
In this review, we applied the tool with eleven questions, and answers with yes, no, somewhat, and can't tell.
We prioritized three questions from the checklist that aligned with the review's objectives, ensuring methodologically rigorous and meaningful information to analyze robot design cues and user perceptions in dementia care.
(1) \textit{Have ethical issues been taken into consideration?}
(2) \textit{Was the data analysis sufficiently rigorous?}
(3) \textit{Is there a clear statement of findings?}
Papers that received a ``No'' as an answer to these three questions were considered not to meet the minimum quality threshold, which led to the exclusion of three papers. 
Thus, the present literature review focuses on 26 papers. The selection pipeline is described in Figure \ref{PRISMA diagram}.

The first author performed the information extraction process of the 26 included papers. For each paper, she recorded the following details: 
(1) \textit{general information}: authors, date of publication, type of paper, reported location of the study; 
(2) \textit{study characteristics}: number of participants, age of participants, gender of participants, dementia severity of participants, type of activity, presence of facilitator, duration of the studies and sessions, robot introduction; 
(3) \textit{social robot information}: name, embodiment, morphology, body parts, surface materials, behaviors; 
(4) \textit{perception and response information}: verbal expression of participants, emotional expression of participants, behaviors of participants.

After extracting paper excerpts describing how PLwD perceived and responded to the robots, we copied and pasted them to a new document and conducted thematic analysis following Braun and Clarke's six-phase approach \cite{braun2006using}.
The first author began by familiarizing herself with the data segments and excerpts from the reviewed papers. She then systematically coded features related to perception and response, based on both behavioral observations (reported by caregivers or researchers) and verbal expressions from PLwD or their caregivers, as reported in the reviewed papers. Relevant data were collated under each code.

Next, two researchers grouped the codes regarding perception and response into potential themes, compiling all associated data for each theme. The remaining two authors then reviewed these preliminary themes. The first author grouped the codes regarding design cues that could lead to potential deception and generated themes. All authors participated in refining the themes, generating clear definitions and names.

Finally, the first author organized and reported the themes in this scoping review. Weekly discussions among the authors were held throughout the process to support iterative reflection on different patterns of how PLwD perceive and respond to social robots and ensure analytical consistency.

\subsection{Authors Positionality}

Before reporting our results, we feel it is important to disclose our stance on PLwD and dementia care. Our perspective aligns with principles of person-centered care \cite{brooker2003person} and warm technology \cite{ijsselsteijnWarmTechnologyNovel2020}, which emphasize the personhood, agency, and lived experience of PLwD. Within this perspective, PLwD are not seen through their cognitive deficits, but are considered as unique individuals, each with abilities, preferences, and capacity for engagement, that may vary from day to day, across situations, and through disease stages.
This perspective shaped the way we examined SRD in this literature review. We interrogated the potentially problematic aspects of the design and perception of social robots with an eye to the agency and autonomy of PLwD.. 
This does not imply a negative stance towards the use of social robots in dementia care. Rather, our intention is to highlight the complexity of their use, and describe the many facets of SRD in dementia care, which intersect with cultural and care practices. 
In this review, SRD is treated as an analytic concept to examine what types of false beliefs are formed in PLwD when they interact with robots (see Section~\ref{sec:results_p&r}), and what contextual (see Sections~\ref{sec:results_framingrobots}) and design cues might be associated with them (see Section~\ref{sec:results_designcues}).

\section{Results}\label{sec:results}

To contextualize the scope of our analysis, Section~\ref{sec:results_includedstudies} and Section~\ref{sec:results_robots} present the relevant information in terms of participants, activities, and social robots.
We report how social robots were introduced and framed by facilitators in Section~\ref{sec:results_framingrobots}. The reported therapeutic benefits for PLwD are summarized in Section~\ref{sec:results_therapeuticoutcomes}. Next, we focus on the design cues built in social robots for PLwD in Section~\ref{sec:results_designcues} and responses of PLwD to social robots that reflect their perceptions in Section~\ref{sec:results_p&r}.

\subsection{Characteristics of the Included Studies}\label{sec:results_includedstudies}

Of the 26 papers included in this review, 19 (73.1\%) are journal papers \cite{abdollahi2022artificial, bradwellImplementingAffordableSocially2022, bradwellUsercenteredDesignCompanion2021, casey2020perceptions, dinesenUseSocialRobot2022, gustafsson2015using, hungExploringPerceptionsPeople2021, inoue2021exploring, kuwamuraCanWeTalk2016, moyle2019using, moyleWhatEffectDoes2016, pikeRobotCompanionCats2021, puHowPeopleDementia2020, rouaixAffectiveEngagementIssues2017,  sarabia2018assistive, sumioka2021minimal, tanioka2021development, whelanEffectsMARIOSocial2020, yamazakiAcceptabilityTeleoperatedAndroid2014}, and seven (26.9\%) are full papers included in the proceedings of a conference \cite{cruz2020social, feng2019livenature, hsu2023co, joshi2019robots, marchetti2022pet, rassInvestigatingPotentialImpacts2023, thunbergRobotPetsDecrease2020}.
According to the reported locations of the studies, the majority were conducted in Europe (n = 14, the UK \cite{bradwellImplementingAffordableSocially2022, bradwellUsercenteredDesignCompanion2021, casey2020perceptions, pikeRobotCompanionCats2021, sarabia2018assistive}, Denmark \cite{dinesenUseSocialRobot2022, marchetti2022pet, yamazakiAcceptabilityTeleoperatedAndroid2014}, Germany \cite{rassInvestigatingPotentialImpacts2023}, Italy \cite{casey2020perceptions}, Ireland \cite{casey2020perceptions, whelanEffectsMARIOSocial2020}, the Netherlands \cite{feng2019livenature}, France \cite{rouaixAffectiveEngagementIssues2017}, and Sweden \cite{gustafsson2015using, thunbergRobotPetsDecrease2020}), followed by America (n = 5, the US \cite{abdollahi2022artificial, hsu2023co, joshi2019robots}, Mexico \cite{cruz2020social}, and Canada \cite{hungExploringPerceptionsPeople2021}), Asia (n = 4, Japan \cite{inoue2021exploring, kuwamuraCanWeTalk2016, sumioka2021minimal, tanioka2021development}), and Oceania (n = 3, Australia \cite{moyle2019using, moyleWhatEffectDoes2016, puHowPeopleDementia2020}), indicating a diverse but still imbalanced representation of the countries and cultures of the participants. 

\subsubsection{Participants}\label{4.1.1}
The 26 studies included a total of 558 PLwD, with sample sizes ranging from 2 to 138 people. The studies included in this review reported age in very different ways (e.g., age ranges, average age).
\setcounter{table}{2}  
\begin{table*}[h]
\centering
\caption{Overview of participant information in the studies included in the scoping review (F = female, M = male, NS = not specified).}
\label{table3_participants}
\begin{tabularx}{\textwidth}{
p{3.5cm}
p{0.8cm}
p{2.5cm}
p{2.5cm}
p{4.5cm}}
\toprule
\textbf{Author} & \textbf{N} & \textbf{Age (M)} & \textbf{Gender} & \textbf{Dementia Severity} \\
\midrule
Abdollahi et al. (2022) & 10 & 77.1 years & 7 F, 3 M & 10 mild \\

Bradwell et al. (2022) & 83 & 87.21 years & 61 F, 22 M & DSRS: $0$--$54$, $M=32.11$ (moderate) \\

Bradwell et al. (2021) & 26 & 62--107 years* & 20 F, 6 M & NS \\

Casey et al. (2020) & 38 &
\makecell[l]{UK: $60+$ years\\Italy: $76+$ years\\Ireland: $70+$, $55+$} &
\makecell[l]{UK: 5 F, 3 M\\ Italy: 12 F, 8 M\\ Ireland: NS} &
\makecell[l]{UK: 8 mild\\Italy: 20 mild\\Ireland: 2 mild, 6 moderate} \\

Cruz-Sandoval et al. (2020) & 9 & 83.77 years & 6 F, 3 M & 2 mild, 7 moderate \\

Dinesen et al. (2022) & 42 &
\makecell[l]{IS: $83$ years\\GS: $84$ years} &
\makecell[l]{IS: 11 F, 1 M\\ GS: 22 F, 8 M} &
42 mild \\

Feng et al. (2019) & 9 & 84.11 years & 7 F, 2 M & 2 mild, 3 moderate, 4 severe \\

Gustafsson et al. (2015) & 4 & 82--90 years & 2 F, 2 M & 4 severe \\

Hsu et al. (2023) & 12 & 66--96 years & 8 F, 4 M & 12 moderate \\

Hung et al. (2021) & 10 & $60+$ years & 4 F, 6 M & 2 mild, 5 moderate, 3 severe \\

Inoue et al. (2021) & 7 & 87.29 years & 6 F, 1 M & 1 mild, 4 moderate, 2 severe \\

Joshi et al. (2019) & 28 & $55+$ years & NS & NS \\

Kuwamura et al. (2016) & 3 & 91.33 years & 3 F & 2 moderate, 1 severe \\

Marchetti et al. (2022) & 30 & NS & 16 F, 14 M & Most are severe \\

Moyle et al. (2019) & 138 & NS & NS & NS \\

Moyle et al. (2016) & 5 & 84 years & 5 F & 2 mild, 3 moderate \\

Pike et al. (2021) & 12 & NS & 11 F, 1 M & Mild, moderate, severe \\

Pu et al. (2020) & 11 & 84.36 years & 9 F, 2 M & MMSE: $9$--$24$ (mild, moderate) \\

Ra\ss{} et al. (2023) & 12 & $64$--$91$ years & 7 F, 5 M & NS \\

Rouaix et al. (2017) & 9 & 86 years & 7 F, 2 M & MMSE: $12$--$22$ (mild, moderate) \\

Sarabia et al. (2018) & 7 & $79$--$99$ years & 7 F & NS \\

Sumioka et al. (2021) & 21 & 86.6 years & 18 F, 3 M & NS \\

Tanioka et al. (2021) & 2 & $90+$ years & NS & 1 moderate, 1 severe \\

Thunberg et al. (2020) & 17 &
\makecell[l]{RC: NS\\RD: $80$--$90$} &
\makecell[l]{RC: NS\\ RD: 1 F, 1 M} &
NS \\

Whelan et al. (2020) & 10 & 83 years & 7 F, 3 M & 2 mild, 6 moderate, 2 severe \\

Yamazaki et al. (2014) & 5 & 87.8 years & 5 F & 5 severe \\
\bottomrule
\end{tabularx}
\end{table*}
As such, we provide information about the age of participants following the format used in each study in Table~\ref{table3_participants}.
The gender of the participants was not always reported in the papers. 
The 23 (88.5\%) papers \cite{abdollahi2022artificial, bradwellImplementingAffordableSocially2022, bradwellUsercenteredDesignCompanion2021, casey2020perceptions, cruz2020social, dinesenUseSocialRobot2022, feng2019livenature, gustafsson2015using, hsu2023co, hungExploringPerceptionsPeople2021, inoue2021exploring, kuwamuraCanWeTalk2016, marchetti2022pet, moyleWhatEffectDoes2016, pikeRobotCompanionCats2021, puHowPeopleDementia2020, rassInvestigatingPotentialImpacts2023, rouaixAffectiveEngagementIssues2017, sarabia2018assistive, sumioka2021minimal, thunbergRobotPetsDecrease2020, whelanEffectsMARIOSocial2020, yamazakiAcceptabilityTeleoperatedAndroid2014} that reported gender show an unequal distribution across participants with 72.8\% women (n = 267) and 27.2\% men (n = 100). This is in line with the highest incidence of dementia in women.
Only 14 (53.8\%) studies \cite{abdollahi2022artificial, casey2020perceptions, cruz2020social, dinesenUseSocialRobot2022, feng2019livenature, gustafsson2015using, hsu2023co, hungExploringPerceptionsPeople2021, inoue2021exploring, kuwamuraCanWeTalk2016, moyleWhatEffectDoes2016, tanioka2021development, whelanEffectsMARIOSocial2020, yamazakiAcceptabilityTeleoperatedAndroid2014} report the dementia severity of participants. Among them, the majority of participants had mild dementia (n = 93, 56.7\%), followed by moderate dementia (n = 49, 29.9\%) and severe dementia (n = 22, 13.4\%).

\subsubsection{Activities}\label{4.1.3}
This section provides details on the duration of the study and session, the type of sessions, and the structure of the activities.

\textit{Study and Session Duration.} 
The length of the included studies ranged from short-term studies, lasting a single session up to a few days, to long-term studies lasting up to nine months. 
Most of the studies fell within the range of 4 to 12 weeks \cite{casey2020perceptions, cruz2020social, dinesenUseSocialRobot2022, feng2019livenature, gustafsson2015using, hsu2023co, hungExploringPerceptionsPeople2021, inoue2021exploring, kuwamuraCanWeTalk2016, moyle2019using, moyleWhatEffectDoes2016, pikeRobotCompanionCats2021, puHowPeopleDementia2020, rassInvestigatingPotentialImpacts2023, thunbergRobotPetsDecrease2020, whelanEffectsMARIOSocial2020}
(See Table~\ref{structures} for more details).

\textit{Type of Sessions.} 
Most studies (n = 15, 57.7\%) \cite{abdollahi2022artificial, casey2020perceptions, feng2019livenature, gustafsson2015using, hungExploringPerceptionsPeople2021, inoue2021exploring, kuwamuraCanWeTalk2016, moyle2019using, moyleWhatEffectDoes2016, pikeRobotCompanionCats2021, rouaixAffectiveEngagementIssues2017, sarabia2018assistive, sumioka2021minimal, whelanEffectsMARIOSocial2020, yamazakiAcceptabilityTeleoperatedAndroid2014} presented individual sessions where only one person with dementia interacted with robots, while fewer studies (n = 8, 30.8\%) \cite{bradwellUsercenteredDesignCompanion2021, cruz2020social, hsu2023co, joshi2019robots, marchetti2022pet, rassInvestigatingPotentialImpacts2023, tanioka2021development, thunbergRobotPetsDecrease2020} focused on group sessions where multiple PLwD interacted with one or more robots simultaneously. Three papers (11.5\%) \cite{bradwellImplementingAffordableSocially2022, dinesenUseSocialRobot2022, puHowPeopleDementia2020} featured both individual and group sessions. 

Of the 26 reviewed papers, only 22 (84.6\%) specified whether or not a facilitator\footnote{someone who assists in activities involving PLwD, either by guiding the interaction or maintaining engagement} was present during the interaction with the robot. In two of these papers (9.1\%) \cite{bradwellUsercenteredDesignCompanion2021, moyle2019using}, PLwD interacted with robots alone. In 20 (90.9\%), they interacted with the robot in the presence of a facilitators, with facilitators being formal caregivers (n = 7, 35.0\%) \cite{dinesenUseSocialRobot2022, gustafsson2015using, hungExploringPerceptionsPeople2021, rassInvestigatingPotentialImpacts2023, rouaixAffectiveEngagementIssues2017, sumioka2021minimal, thunbergRobotPetsDecrease2020}, informal caregivers (n = 2, 10.0\%) \cite{inoue2021exploring, pikeRobotCompanionCats2021}, researchers (n = 8, 40.0\%) \cite{casey2020perceptions, feng2019livenature, hsu2023co, marchetti2022pet, moyleWhatEffectDoes2016, sarabia2018assistive, whelanEffectsMARIOSocial2020, yamazakiAcceptabilityTeleoperatedAndroid2014}, students (n = 1, 5.0\%) \cite{kuwamuraCanWeTalk2016}, or a combination of formal caregiver and a researcher (n = 2, 10.0\%) \cite{joshi2019robots, tanioka2021development}.

\textit{Types of Activities.} 
Most of the articles (n = 21, 80.8\%) \cite{bradwellImplementingAffordableSocially2022, bradwellUsercenteredDesignCompanion2021, casey2020perceptions, dinesenUseSocialRobot2022, feng2019livenature, gustafsson2015using, hungExploringPerceptionsPeople2021, inoue2021exploring, joshi2019robots, kuwamuraCanWeTalk2016, marchetti2022pet, moyle2019using, moyleWhatEffectDoes2016, pikeRobotCompanionCats2021, puHowPeopleDementia2020, sarabia2018assistive, sumioka2021minimal, tanioka2021development, thunbergRobotPetsDecrease2020, whelanEffectsMARIOSocial2020, yamazakiAcceptabilityTeleoperatedAndroid2014} focused on free play with the robot, the robots being left with PLwD, and the PLwD having free interaction with the robot.
The remaining papers (n = 5, 19.2\%) \cite{abdollahi2022artificial, cruz2020social, hsu2023co, rassInvestigatingPotentialImpacts2023, rouaixAffectiveEngagementIssues2017} focused on structured activities in which robots are brought to PLwD to engage them in specific interactions.
The activities range from physical activities to cognitive stimulation sessions, as shown in Table~\ref{activities-only}.

\subsection{Framing the Nature of Social Robots}\label{sec:results_framingrobots}
In this section, we present the different ways in which facilitators introduce social robots to PLwD in the reviewed papers. 

\textit{Introducing the robot.} As shown in Table~\ref{Introduction}, out of the 26 reviewed papers, only seven (26.9\%) provided descriptions of how robots were first introduced to PLwD. Among these, four (57.1\%) studies introduced robots with clear identification as robotic entities, such as saying ``\textit{This is a social robot}'' and explaining how it interacts with people \cite{gustafsson2015using,hungExploringPerceptionsPeople2021,moyleWhatEffectDoes2016,yamazakiAcceptabilityTeleoperatedAndroid2014}, while in one study (14.3\%), the caregiver introduced the robot similarly to a human baby \cite{sumioka2021minimal} (see Tables \ref{Introduction} and \ref{framing}). Two studies (28.6\%) \cite{marchetti2022pet, joshi2019robots} adopted a neutral approach. In these cases, the facilitators did not clearly define the robot's ontological status. Instead, the robots were referred to simply as ``\textit{it,}'' or participants were invited to form their own interpretations (See Table~\ref{Introduction}).

\textit{Frame nature during interaction} As shown in Table~\ref{framing}, in five (38.5\%) out of 13 studies that reported facilitators' framing, the facilitators maintained the illusion of lifelikeness, mainly by interpreting the robot's behavioral cues such as breathing \cite{gustafsson2015using} or baby voice \cite{kuwamuraCanWeTalk2016}, intentions such as going to live with the person with dementia or trying to say something \cite{marchetti2022pet, inoue2021exploring}, or emotions such as loving the person with dementia \cite{hungExploringPerceptionsPeople2021}. 
In three (23.1\%) papers, facilitators used prompts or mentioned beneficial outcomes to motivate PLwD to interact with social robots \cite{bradwellImplementingAffordableSocially2022,pikeRobotCompanionCats2021,sumioka2021minimal}. For instance, motivating a person with dementia to get up in the morning by saying JfA Cat was waiting for the person \cite{pikeRobotCompanionCats2021}.
In two (15.4\%) papers, facilitators responded to PLwD's questions about the nature of the robot by redirecting the conversation \cite{inoue2021exploring,marchetti2022pet}. While a more neutral or truthful approach was observed in eight (61.5\%) studies, where facilitators explained the robot's nature with clear information about their artifactuality and functionalities \cite{gustafsson2015using,feng2019livenature,hungExploringPerceptionsPeople2021,joshi2019robots,marchetti2022pet,moyleWhatEffectDoes2016,rassInvestigatingPotentialImpacts2023,thunbergRobotPetsDecrease2020}. 
For example, a care staff explained \cite{hungExploringPerceptionsPeople2021}, 
\begin{quote}
    \textit{This is a social robot, PARO; it has sensors in his whiskers. If you talk to it, it responds to you.}
\end{quote}

\textit{Debriefing.} None of the 26 included studies explicitly reported a debriefing process. It remains unclear whether debriefing was omitted in the paper or simply not conducted.

\begin{table*}[p]
\centering
\rotatebox{90}{%
  \begin{minipage}{\textheight}
    \centering
    \caption{Overview of interaction characteristics, robot design features, and herapeutic outcomes in the studies included in the scoping review (Abbreviations: CLP = Co-located physical, PR = Partially remote, AM = Anthropomorphic, BM = Biomorphic, AB = Abstract, FB = Full-body, PB = Partial-body, HP = Hard plastic and metal, SF = Soft fur or plush fabrics, SS = Semantic speech, SFS = Semantic-free speech, NS = not specified).}
    \label{table4_interaction_design_outcomes}

    \adjustbox{max width=\textheight}{%
      \begin{tabularx}{\textheight}{
      p{3.5cm}
      p{2cm}
      p{1.2cm}
      p{1.2cm}
      p{1.2cm}
      p{1.2cm}
      p{1.2cm}
      X}
      \toprule
      \textbf{Author} &
      \textbf{Robot} &
      \textbf{Embodiment} &
      \textbf{Form} &
      \textbf{Body structure} &
      \textbf{Surface material} &
      \textbf{Sound / speech} &
      \\textbf{Therapeutic outcomes} \\
      \midrule

      Abdollahi et al. (2022) & Ryan & CLP & AM & PB & HP & SS &
      mood improvement; positive user experience; communication engagement; social isolation reduction; depression reduction* (PHQ-9 \& GDS); individual variability \\

      Bradwell et al. (2022) & JfA Cat, JfA Dog & CLP & BM & FB & SF & SFS &
      communication and daily activity engagement; isolation and loneliness reduction; neuropsychiatric symptom reduction (NPI)*; individual variability \\

      Bradwell et al. (2021) & Pleo, Miro, PP Dog, Paro, JfA Dog, JfA Cat, Furby, Hedgehog & CLP & BM & FB & HP & SFS &
      communication engagement; robot variability \\

      Casey et al. (2020) & Mario & CLP & AM & FB & HP & SS &
      mood improvement; positive user experience; quality of life improvement; autonomy enhancement; communication engagement; loneliness reduction \\

      Cruz-Sandoval et al. (2020) & Eva & CLP & AM & FB & HP & SS &
      mood improvement; daily activity engagement; socialization stimulation; reduction of delusions, agitation/aggression, and euphoria/exaltation** (NPI-NH); individual variability \\

      Dinesen et al. (2022) & LOVOT & CLP & BM & FB & HP & SFS &
      mood improvement; well-being improvement* (WHO-5); communication engagement \\

      Feng et al. (2019) & Pleo & CLP & BM & FB & HP & SFS &
      alertness improvement** and pleasure increase* (OERS); communication facilitation; memory recollection \\

      Gustafsson et al. (2015) & JustoCat & CLP & BM & FB & SF & SFS &
      quality of life improvement; well-being improvement* (QUALID); communication and activity engagement; loneliness reduction; agitation reduction* (CMAI); memory recollection \\

      Hsu et al. (2023) & QT & CLP & AM & FB & HP & SS &
      overall well-being increase; autonomy improvement; social security and connectedness improvement \\

      Hung et al. (2021) & Paro & CLP & BM & FB & SF & SFS &
      positive emotion; comfort, self, and safety; social connection facilitation \\

      Inoue et al. (2021) & Paro & CLP & BM & FB & SF & SFS &
      comfort; well-being improvement; fulfillment of needs for occupation, identity, and inclusion; behavioral improvement in articulation, reminiscence, intellectual engagement, attachment* (BCC) \\

      Joshi et al. (2019) & Paro, JfA Cat, Nao, Cozmo & CLP & BM \& AM \& AB & FB \& PB & SF \& HP & SFS \& SS &
      communication and collaboration engagement; social connection facilitation; cognitive stimulation through memory recall \\

      Kuwamura et al. (2016) & Telenoid & PR & AM & PB & HP & NS &
      communication improvement** (questionnaire, Q1-Q3); individual variability \\

      Marchetti et al. (2022) & Sanne & CLP & BM & FB & HP & NS &
      communication facilitation \\

      Moyle et al. (2019) & Paro & CLP & BM & FB & SF & SFS &
      individual variability \\

      Moyle et al. (2016) & CuDDler & CLP & BM & FB & SF & SFS &
      agitation reduction* (CMAI); individual variability \\

      Pike et al. (2021) & JfA Cat & CLP & BM & FB & SF & SFS &
      positive emotion; calming and anxiety reduction; conversation and daily activity stimulation; distraction from repetitive behaviors; memory improvement \\
      
      \bottomrule
      \end{tabularx}
    }
  \end{minipage}%
}
\end{table*}

\begin{table*}[p]
\ContinuedFloat
\rotatebox{90}{%
  \begin{minipage}{\textheight}

    \caption{Overview of interaction characteristics, robot design features, and therapeutic outcomes in the studies included in the scoping review (continued).}

    \adjustbox{max width=\textheight}{%
      \begin{tabularx}{\textheight}{
      p{3.5cm}
      p{2cm}
      p{1.2cm}
      p{1.2cm}
      p{1.2cm}
      p{1.2cm}
      p{1.2cm}
      X}
      \toprule
      \textbf{Author} &
      \textbf{Robot} &
      \textbf{Embod.} &
      \textbf{Form} &
      \textbf{Body structure} &
      \textbf{Surface material} &
      \textbf{Sound / speech} &
      \textbf{Therapeutic outcomes} \\
      \midrule
      
      Pu et al. (2020) & Paro & CLP & BM & FB & SF & SFS &
      mood improvement; pain relief; sleep promotion; individual variability; loneliness reduction; positive memory recall \\

      Ra\ss{} et al. (2023) & Pepper & CLP & AM & FB & HP & SS &
      physical activity activation; engagement in group sessions and communication; memory reactivation \\

      Rouaix et al. (2017) & Nao & CLP & AM & FB & HP & SS &
      positive affect increase* (PANAS); pleasure-related response increase** (MPES); well-being improvement* \\

      Sarabia et al. (2018) & Nao & CLP & AM & FB & HP & SS &
      mood improvement; intra-individual fluctuating emotional responses; physical activity engagement \\

      Sumioka et al. (2021) & Hiro & CLP & AM & FB & SF & SFS &
      positive user experience and calmness; individual variability \\

      Tanioka et al. (2021) & Pepper & PR & AM & FB & HP & SS & 
      \\

      Thunberg et al. (2020) & JfA Cat, JfA Dog & CLP & BM & FB & SF & SFS &
      individual and robot variability; conversation facilitation; long-term calmness; agitation reduction \\

      Whelan et al. (2020) & Mario & CLP & AM & FB & HP & SS &
      positive mood and mean attitudes* (OME); mood elevation; increased opportunities for meaningful activities \\

      Yamazaki et al. (2014) & Telenoid & PR & AM & PB & HP & NS &
      communication encouragement \\

      \bottomrule
      \end{tabularx}
    }
    \vspace{3mm}

    \footnotesize
    \textit{Notes:}
    * Quantitative assessment.
    ** Statistically significant result.
    
    Abbreviations: PHQ-9 = Patient Health Questionnaire-9; 
    GDS = Geriatric Depression Scale; 
    NPI = Neuropsychiatric Inventory; 
    NPI-NH = Japanese version of the Neuropsychiatric Inventory; 
    WHO-5 = World Health Organization-Five Well-Being Index;
    OERS = Observed Emotion Rating Scale; 
    QUALID = Quality of Life in Late-Stage Dementia scale; 
    CMAI = Cohen–Mansfield Agitation Inventory; 
    PANAS = International Positive and Negative Affect Schedule; 
    MPES = Menorah Park Engagement Scale; 
    OME = Observed Mood and Engagement scale; 
    BCC = Behavioral Coding Category.
  \end{minipage}%
}
\end{table*}

\subsection{Social Robots}\label{sec:results_robots}
 
The included studies employed 21 robots (see Figure~\ref{Robots}). Four studies \cite{bradwellImplementingAffordableSocially2022, bradwellUsercenteredDesignCompanion2021, joshi2019robots, thunbergRobotPetsDecrease2020} involved multiple robots, while the others focused on a single robot.
In this section, we describe the robots included in these reviewed studies, focusing on their embodiment, morphology, body structure, surface material, and sound/speech (see Table~\ref{table4_interaction_design_outcomes}).

\setcounter{table}{4}  
\begin{table*}[h]
\caption{Introduction of the robots in the included studies.}
\label{Introduction}
\centering
\begin{tabularx}{\textwidth}{>{\arraybackslash}p{3.5cm} >{\arraybackslash}X}
\toprule
Author (Date) & Description of the introduction of the robots \\
\midrule

Gustafsson et al. (2015) \cite{gustafsson2015using}& The professional caregivers presented JustoCat and demonstrated how to stroke it and make it purr. In approaching participants with JustoCat, it was presented as a robotic pet and not a live animal. \\
Hung et al. (2021) \cite{hungExploringPerceptionsPeople2021}& Staff Gail: (Showing PARO to Max) "This is a social robot, PARO; it has sensors in his whiskers. If you talk to it, it responds to you." \\
Yamazaki et al. (2014) \cite{yamazakiAcceptabilityTeleoperatedAndroid2014}& Prior to the trial, one of our staff and caretakers introduced Telenoid to the participants by showing its picture and explaining that it was a robot for communication, and that somebody would be on the other end and communicate through the robot. Immediately before the session, we put Telenoid on the stand in front of the participants and explained again that somebody would operate the robot from another room and communicate through it. \\
Moyle et al. (2016) \cite{moyleWhatEffectDoes2016}& CuDDler was introduced to each participant with a statement of “Hello participant. This is CuDDler. CuDDLer is a robotic bear. Would you like to get to know CuDDler?” \\
\midrule
Joshi et al. (2019) \cite{joshi2019robots}& We showed them slides about social robots used for social and assistive interactions with older adults and children and discussed their benefits in dementia care. For one session using Paro, the preschool teacher sourced a library book and conducted a storytelling activity about `baby seals' before introducing the robot. Two Paro robots were then placed on the center table in the living room, and children and residents were asked what they thought it was, whether it was a living thing, a baby seal, or a robot. \\
Marchetti et al. (2022) \cite{marchetti2022pet}& Sanne was introduced 7 times as a floor cleaner/washer, 6 times as a cat, and in 5 situations, she was neutrally referred to as ``it'. \\
\midrule
Sumioka et al. (2021) \cite{sumioka2021minimal}& The staff and an experimenter entered the participant's room, introduced the robot they were holding. The staff and experimenter treated the robot like a human baby. \\

\bottomrule
\end{tabularx}
\end{table*}

\begin{table*}[htbp]
\caption{Framing nature of the robots during interaction.}
\label{framing}
\centering
\begin{tabularx}{\textwidth}{>{\arraybackslash}p{3.5cm} >{\arraybackslash}X}
\toprule
Author (Date) & Description of the framing of the robots nature \\
\midrule
Gustafsson et al. (2015) \cite{gustafsson2015using}&  Is it smooth? Is it breathing? Is it purring?  \\
Kuwamura et al. (2016) \cite{kuwamuraCanWeTalk2016}& The speakers also often adapted to the participants by changing their voice using a voice changer to sound more like a child. \\
Marchetti et al. (2022) \cite{marchetti2022pet}& It's going to live here, It's Sanne; Look, the cat is coming there. \\
Inoue et al. (2021) \cite{inoue2021exploring}& PARO has arrived; PARO is saying hello; PARO is looking at you. \\
Hung et al. (2021) \cite{hungExploringPerceptionsPeople2021}& I think it [PARO] likes you. It's looking right at you, Max. What do you think it's trying to say? \\
\midrule
Bradwell et al. (2022) \cite{bradwellImplementingAffordableSocially2022}& Oh, can you just keep an eye on the dog (or the puppy)?  \\
Pike et al. (2021) \cite{pikeRobotCompanionCats2021}& (The cat's name) is waiting for you (to prompt PLwD to get out of bed in the morning). \\
Sumioka et al. (2021) \cite{sumioka2021minimal}& He (the experimenter) needs my help. Do you mind looking after this baby? \\
\midrule
Inoue et al. (2021) \cite{inoue2021exploring}& My mother (a person with dementia) asked, ``Is this child a seal?'' So, I responded to her with another question, ``Do you know where this child (PARO) came from?'' My mother answered, ``I wonder maybe somewhere cold?'' \\
Marchetti et al. (2022) \cite{marchetti2022pet}& H (a person with dementia): ``Hello little dog.'' H reaches out a hand, bends over and softly pets Sanne's nose: ``Yeah that's nice.'' She changes from stroking with the back of her hand to more intense touching of head and ears. H: ``Uh it's hard on top.'' Staff: ``Doesn't it look nice?'' H: ``It looks really cute and could easily be a regular one. But it can't say something.'' \\
\bottomrule
\end{tabularx}
\end{table*}

\textit{Embodiment.}
Based on the classification of Wainer et al. \cite{wainer2006role}, all robots used in the reviewed studies were co-located physical robots.
Only Telenoid in \cite{kuwamuraCanWeTalk2016, yamazakiAcceptabilityTeleoperatedAndroid2014} was remotely teleoperated, and Pepper in \cite{tanioka2021development} was controlled by operators with a Wizard-of-Oz technique (see Figure~\ref{Robots}).

\textit{Form.}
Based on the classification by Bartneck \& Forlizzi \cite{bartneck2004design}, the majority of robots used in the reviewed studies had a biomorphic form (n = 12, 57.1\%), followed by the anthropomorphic form (n = 8, 38.1\%). Only one robot (4.8\%), Cozmo, had an abstract form.
Most of the biomorphic robots resembled existing animals such as seals, cats, and dogs (see Figure~\ref{Robots}).

\textit{Body structure.} 
Most robots in the studies (n = 18, 90.0\%) had a full-body design. A small portion of them (n = 2, 10.0\%) featured only the upper body, such as the head and torso (Ryan and Telenoid). Cozmo was not included because of its abstract form (see Figure~\ref{Robots}).

\textit{Surface material.} \label{4.1.2m}
Most robots (n = 13, 61.9\%) were made of hard plastic and metal materials, while a smaller percentage (n = 8, 38.1\%) of the robots were made of soft materials such as synthetic fur and plushy fabric.
Most anthropomorphic robots (n = 7, 87.5\%) were made of hard materials except for Hiro, and most zoomorphic robots (n = 8, 66.7\%) were made of soft materials except for Pleo, Miro, Hedgehog, and Sanne.

\textit{Sounds/Speech.} \label{4.1.2v}
Following the classification in Robinson et al. \cite{robinson2022designing}, most robots (n = 11, 52.3\%) in the included studies could make nonverbal sound and semantic-free speech that does not convey specific meaning, such as pet sounds and baby crying, while a smaller portion of the robots (n = 6, 28.6\%) in the included studies could verbally communicate, and an even smaller fraction of these robots (n = 4, 19.0\%) did not produce sound at all. 

\subsection{Therapeutic Outcomes}\label{sec:results_therapeuticoutcomes}
Across the 26 included studies, therapeutic outcomes were predominantly psychosocial, behavioral, and cognitive.

Most studies descriptively reported psychosocial changes, including improvements in mood or positive affect, such as calm, enjoyment, comfort, and relief from stress, anxiety, and agitation \cite{abdollahi2022artificial, bradwellImplementingAffordableSocially2022, casey2020perceptions, cruz2020social, dinesenUseSocialRobot2022, feng2019livenature, hsu2023co, hungExploringPerceptionsPeople2021, inoue2021exploring, moyleWhatEffectDoes2016, pikeRobotCompanionCats2021, puHowPeopleDementia2020, rouaixAffectiveEngagementIssues2017, sarabia2018assistive, sumioka2021minimal, whelanEffectsMARIOSocial2020, thunbergRobotPetsDecrease2020}. 
Many studies further reported reduced loneliness, facilitated social connectedness and inclusion as one of the outcomes of interacting with social robots \cite{abdollahi2022artificial, bradwellImplementingAffordableSocially2022, casey2020perceptions, cruz2020social, gustafsson2015using, hsu2023co, hungExploringPerceptionsPeople2021, inoue2021exploring, joshi2019robots, puHowPeopleDementia2020}.
Broader psychosocial benefits, including enhanced well-being or quality of life, sense of self, safety and security, and autonomy, were also described \cite{casey2020perceptions, gustafsson2015using, hsu2023co, inoue2021exploring, rouaixAffectiveEngagementIssues2017, hsu2023co, hungExploringPerceptionsPeople2021}.

Only nine studies quantitatively assessed the changes. Rouaix et al. reported that interacting with social robots significantly increased pleasure-related responses \cite{rouaixAffectiveEngagementIssues2017}, Cruz-Sandoval et al. found a statistically significant reduction of delusions, agitation/aggresion, and euphoria/exaltation \cite{cruz2020social}, and Feng et al. reported that alertness was significantly increased \cite{feng2019livenature}.
Dinensen et al., Rouaix et al., and Gustaffson et al. reported a non-significant increase in well-being \cite{dinesenUseSocialRobot2022, rouaixAffectiveEngagementIssues2017, gustafsson2015using} and Abdollahi et al., Bradwell et al., and Cruz et al. a non-significant decrease in neuropsychiatric symptoms \cite{abdollahi2022artificial,bradwellImplementingAffordableSocially2022, cruz2020social}.

Behavioral changes were frequently reported, including increased engagement in communication with others and participation in meaningful activities \cite{abdollahi2022artificial, bradwellImplementingAffordableSocially2022, bradwellUsercenteredDesignCompanion2021, casey2020perceptions, cruz2020social, dinesenUseSocialRobot2022, feng2019livenature, gustafsson2015using, joshi2019robots, marchetti2022pet, pikeRobotCompanionCats2021, rassInvestigatingPotentialImpacts2023, sarabia2018assistive, thunbergRobotPetsDecrease2020, whelanEffectsMARIOSocial2020, yamazakiAcceptabilityTeleoperatedAndroid2014}, or improved behaviors such as articulation and use of intellectual abilities \cite{inoue2021exploring}. Only one study reporting improved communication reached statistical significance \cite{kuwamuraCanWeTalk2016}, while two studies quantitatively evaluated a reduction in agitation \cite{gustafsson2015using, moyleWhatEffectDoes2016} and one study on behavioral improvement \cite{inoue2021exploring} did not reach statistical significance.

Some studies reported that interaction with social robots evoked memories among PLwD, suggesting cognitive stimulation. \cite{feng2019livenature, gustafsson2015using, joshi2019robots, pikeRobotCompanionCats2021, puHowPeopleDementia2020, rassInvestigatingPotentialImpacts2023}.

\subsection{Design Cues Embedded in Social Robots} \label{sec:results_designcues}
In this section, we present the design cues embedded in social robots as reported in the reviewed studies. 
The aim here is descriptive: systematically identifying and organizing the design and interaction cues used in social robots for PLwD in the HRI literature. This is a necessary analytical step that will allow us to take a more critical stance and discuss which of these cues might lead to SRD in PLwD in Section \ref{sec:discussion}.
The first author identified and compiled all design cues and organized them into four categories. These categories function as interpretative lenses rather than strict or mutually exclusive classifications, and do not necessarily lead to SRD.

\subsubsection{Design Cues Resembling Physiological Signs} 

Several studies incorporated subtle movements to mimic physiological signs of life, signaling that the robot is alive, even when the robot was not actively interacting.
For example, Rouaix et al. \cite{rouaixAffectiveEngagementIssues2017} programmed Nao to slightly undulate when not talking, giving the impression of breathing and being alive. 
The simulated heartbeat rhythm was included as a design feature in the robots used in the studies by Bradwell et al. \cite{bradwellUsercenteredDesignCompanion2021} and Thunberg et al.\cite{thunbergRobotPetsDecrease2020}.
Blinking was another movement reported among the included studies \cite{moyleWhatEffectDoes2016, inoue2021exploring, bradwellUsercenteredDesignCompanion2021}. Biological needs, such as hunger, can be found in Ryan's speech, saying ``\textit{I (Ryan) sure am feeling hungry now}'' \cite{abdollahi2022artificial}.

\subsubsection{Design Cues Resembling Social Intentions}

Some cues could suggest social intentions.
For example, Bradwell et al. \cite{bradwellImplementingAffordableSocially2022} described how the JfA Cat and Dog performed playful movements such as rolling on their back to invite interaction, like a real dog or cat.
In addition to nonverbal cues, 
Nao's speech in Rouaix et al. \cite{rouaixAffectiveEngagementIssues2017} was designed to make a ``well-mannered'' contact with participants by asking about their feelings.

\subsubsection{Design Cues Resembling Familiar Beings}

Most robot designs displayed combinations of movements mimicking humans or familiar animals, as reported in Section~\ref{sec:results_robots}. 
For example, Bradwell et al. \cite{bradwellImplementingAffordableSocially2022, bradwellUsercenteredDesignCompanion2021} reported JfA Cat and Dog featured body and head movements such as turning the head, lifting the head up, and rolling over onto the back to be tickled like real cats and dogs. In addition, Paro exhibited more specific body and head movements similar to seals, such as wagging its back flippers \cite{hungExploringPerceptionsPeople2021}.
Rouaix et al. \cite{rouaixAffectiveEngagementIssues2017} designed Nao to perform some childlike gestures to align with its size and childlike appearance.
In terms of gender, Abdollahi et al. \cite{abdollahi2022artificial} adopted female characteristics such as the appearance of a 3D animated face and a female voice.

Semantic-free speech related to similar animal/human was widely used. Fourteen (53.8\%) studies deployed social robots that make semantic-free speech, such as meowing and purring of cat-like robots \cite{bradwellImplementingAffordableSocially2022, bradwellUsercenteredDesignCompanion2021, gustafsson2015using, joshi2019robots, pikeRobotCompanionCats2021, thunbergRobotPetsDecrease2020}, barking of dog-like robots \cite{bradwellImplementingAffordableSocially2022, bradwellUsercenteredDesignCompanion2021, thunbergRobotPetsDecrease2020}, cooing of seal-like robots \cite{bradwellUsercenteredDesignCompanion2021, hungExploringPerceptionsPeople2021, inoue2021exploring, joshi2019robots, moyle2019using, puHowPeopleDementia2020}, and crying of baby-like robot \cite{sumioka2021minimal}. 


\subsubsection{Design Cues Revealing Artificial Nature} \label{4.3.2}
While some design cues might enhance lifelike qualities, others might signal the robot's artificial nature. 
Hsu et al. \cite{hsu2023co} reported that the body movements of the QT robot did not meet the expectations of PLwD of dance abilities because they were too primitive due to technological limitations.
In other instances, the robot's design appeared to intentionally preserve aspects of its artificial nature. CuDDler and Furby, for example, retained simplified toy-like appearances with plush textures \cite{moyleWhatEffectDoes2016, bradwellUsercenteredDesignCompanion2021}. 
Robots with plastic materials (as reported in Sec~\ref{4.1.2m}) also reflect the design choices that do not prioritize visual mimicry of humans or animals.


\subsection{Perceptions and Responses of PLwD Towards Social Robots}\label{sec:results_p&r}

In this subsection, we look at how PLwD perceives and responds to social robots during interaction, and report six main themes that emerged from the thematic analysis (see process in Section ~\ref{sec:method}).
The following sub-sections present the themes and sub-themes in detail. 

\subsubsection{PLwD Attribute Biological Traits to Social Robots}
One prominent theme emerging from the analysis highlights how PLwD attribute biological traits to social robots by associating operational states, movement behaviors, and physical characteristics with signs of life.  
For example, Bradwell et al. \cite{bradwellUsercenteredDesignCompanion2021} reported that PP Dog was perceived as ``\textit{dead, poor old sod}'' because of noninteractive movement, and PLwD commented that Miro may be ``\textit{sick}'' when it was turned off.  
Pepper was placed in sleep mode, leading a resident to interpret, ``\textit{He is so saggy, isn't he?}'' \cite{rassInvestigatingPotentialImpacts2023}.  
Thunberg et al. \cite{thunbergRobotPetsDecrease2020} also reported a person with dementia asking if JfA Cat was injured because ``\textit{she had never seen the cat jump or walk.}'' They further observed PLwD interpreting the hard texture of JfA Cat as injuries, saying, ``\textit{It must be injured.}''

PLwD have also been observed to attribute biological feelings, such as hunger or pain, to social robots.  
Sumioka et al. \cite{sumioka2021minimal} reported that a crying sound was perceived as an indication of hunger.  
Similarly, some PLwD interpreted the meowing and barking sounds of JfA Cat and Dog, as well as rolling over on their back, as signals of pain or hunger \cite{thunbergRobotPetsDecrease2020}. 
Although no further explanation was provided, Ryan was perceived as not hungry by PLwD \cite{abdollahi2022artificial}:  
\begin{quote}  
\textit{I would like to take her out to dinner, but she wasn't hungry. Maybe next time.}  
\end{quote}

\subsubsection{PLwD Attribute Social Categories to Social Robots}

Gender and age are the attributed social categories seen in the reviewed studies.  
Although in most cases gender attribution is only observed through verbal references by PLwD without further reasoning, Abdollahi et al. \cite{abdollahi2022artificial} reported that some PLwD referred to the Ryan robot with feminine pronouns due to its face and voice.  
Several studies have reported that PLwD often interact with robots such as Telenoid, Hiro, and LOVOT by changing voice tones and physically engaging as if they were children or infants \cite{bradwellUsercenteredDesignCompanion2021,dinesenUseSocialRobot2022,inoue2021exploring,kuwamuraCanWeTalk2016}, showing that age attribution extends beyond embodiment, body structure, surface material, and speech ability.  
A person with dementia even described a perceived age more specifically, saying  \cite{rassInvestigatingPotentialImpacts2023}, 

\begin{quote}
    \textit{He must be eleven or twelve years old.} 
\end{quote}

\subsubsection{PLwD Perceive Social Robots as Having Mental Capacities}

The third theme reflects PLwD perceiving social robots as having mental capabilities. Thunberg et al. \cite{thunbergRobotPetsDecrease2020} study report that PLwD generally perceive the purring sound of zoomorphic robots as a signal of calmness, as seen in their statement, 

\begin{quote}
    \textit{At first when he (JfA Cat) is with me, he is a bit worried but after a while, he calms down, that is so cozy.}
\end{quote}  

PLwD project human-like intentionality onto robots, interpreting their actions as deliberate and meaningful. For example, Sarabia et al. \cite{sarabia2018assistive} report PLwD perceived Hiro's babbling as denying, saying, ``\textit{Don't say no,}'' while others perceived it as an introduction, stating, ``\textit{It said its name is Kentaro}''. Hsu et al. \cite{hsu2023co} report two PLwD discussed QT's intention behind blinking, interpreting it as: 

\begin{quote}
\textit{– He has a lot to say...I think he's winking at me.}

\textit{– Oh, no, he's blinking at me.}
\end{quote}

\begin{quote}
\textit{– We're new to what he is expecting from us. And he's getting acquainted with it. So I think he's happy to do that.}

\textit{–...When he acts jolly and friendly, then he expects people to return that, I think.}
\end{quote} 

Bradwell et al. \cite{bradwellUsercenteredDesignCompanion2021} report similar discussions around the intention of laughing occurred with Furby:

\begin{quote}
    \textit{– He's laughing because I'm tickling his belly.}
    
    \textit{– Oh, I thought he was laughing at your face!}
\end{quote}

Similarly, Hung et al. \cite{hungExploringPerceptionsPeople2021} report PLwD interpreted Paro's cooing, head movements, gaze, blinking, and specific seal-like motions such as wagging its flippers as signals of the robot showing love and interest toward them.  

\begin{quote}
    \textit{PARO: (Moved its head and returned its gaze to Max with wide opened eyes. Then, it cooed with a nod.)} 
    
   \textit{ Max: (Max smiled). Oh man, you like me? }
    
   \textit{ PARO: (leaned its body on Max, turned its head down and wagged its back flippers) }
    
    \textit{Max: What? I thought you liked me? (laughed) }
    
    \textit{PARO: (cooed, moved its head, and looked at Max; stayed still for a minute) }
    
    \textit{Max: Yes, you like me. (Cheered, raised his palm in the air.) Good, give me ten.}
\end{quote}

\subsubsection{PLwD Assign Social Roles to Social Robots}

The fourth theme emphasizes how PLwD actively define the social roles of social robots, treating them as if they were pets, children, students, or other companion figures from their familiar social relationships. These roles reflect how people make sense of the robot through recognizable patterns of human-animal or human-human interaction.

PLwD often adopt caregiving or ownership roles toward social robots. For instance, Dinesen et al. \cite{dinesenUseSocialRobot2022} report a caregiver observed a person with dementia ``\textit{stepping into a mother role}'' with LOVOT, treating it like a child. Healthcare professionals noted in a focus group, ``\textit{She sits and rocks her leg just like you do with an infant or at least a little baby. She really just wants to sit with it and then just have that feeling}''.

Hung et al. \cite{hungExploringPerceptionsPeople2021} reported a person with dementia referring to Paro as ``\textit{This is cute, my pet. I like him.}'' Similarly, PP Dog, JfA Cat, and JfA Dog were frequently mentioned as being adopted by residents. Thunberg et al. \cite{thunbergRobotPetsDecrease2020} noted that three of the resident women in departments A and B ``\textit{adopted}'' JfA Cat on the same day, believing it was their own cat. 

Yamazaki et al. \cite{yamazakiAcceptabilityTeleoperatedAndroid2014} documented a case where a person with dementia viewed Telenoid as a ``\textit{student.}'' As the conversation progressed, the participant took books from the shelf and started discussing literature with Telenoid, sometimes as if it were his student.  

PLwD also treated robots as meaningful companions. For example, 
Raß et al. \cite{rassInvestigatingPotentialImpacts2023} described PLwD who asked for Pepper's name during the first encounter and showed reluctance to part from Pepper after a short session.
Taniko et al. \cite{tanioka2021development} reported that PLwD placed JfA Dog in a walker or in bed as a companion, and described what was happening on TV to it while watching.

The social role attribution also shows dynamism. Marchetti et al. \cite{marchetti2022pet} observed that a participant sometimes gave Sanne commands like ``\textit{go over there}'' and displayed either relaxation or frustration depending on Sanne's obedience. At other times, she spoke to Sanne in a friendly and affectionate voice, referred to herself as ``\textit{Mama,}'' and asked Sanne to come closer.  

\begin{quote}
\textit{G (in a low voice): ``Go over there. There, over there.'' (G looks a little angry. Sanne wiggles on the spot) G: ``Will you not listen to what I'm saying? You need to do what I'm saying. – Now!''}  
\end{quote}  

\begin{quote}
\textit{G: ``Sanne! Go to Mama. Just gallop a little, friend!'' (Sanne comes closer) G: ``Yeah, that's fine. You have to come here!'' (Sanne drives backwards again) G: ``No, not that way.'' (... some back and forth of the robot, conversation with care staff ...) G (high voice): ``Go over there now. To little Mom. Come over here now, it must be now.'' (Sanne drives away) G: ``Now you must go. Move along now. Away with you. Go back. Yes, you have to go home.''}  
\end{quote}

\subsubsection{PLwD Show Empathy and Caretaking Tendency toward Social Robots}
The fifth theme is that PLwD express empathy toward social robots, perceiving them as entities requiring care and attention. PLwD often display emotional concern for social robots. For example, Thunberg et al. \cite{thunbergRobotPetsDecrease2020} report a person with dementia screamed and looked for JfA Dog when it was taken away for cleaning and changing battery. Similarly, Bradwell et al. \cite{bradwellUsercenteredDesignCompanion2021} reported that a person with dementia demonstrated empathy by gently promising not to harm the JfA Dog, saying, ``\textit{I won't hurt you, darling.}'' Moreover, Marchetti et al. \cite{marchetti2022pet} reported that a participant expressed empathy towards the robot Sanne by remarking:

\begin{quote}
    \textit{I feel bad for you that you have to be inside a shell.}
\end{quote}  

Bradwell et al. \cite{bradwellImplementingAffordableSocially2022} also mentioned a person with dementia, despite realizing the robotic nature of JfA Cat, expressed empathy by commenting ``\textit{the poor cat has got two broken legs. Good job it's not real!}'' 

Further, PLwD often exhibit a strong sense of care toward social robots, perceiving them as helpless creatures requiring care and affection. Pike et al. \cite{pikeRobotCompanionCats2021} reported that participants frequently inquired about feeding or letting JfA Cat out. Bradwell et al. noted a person with dementia enjoyed feeding a JfA Cat \cite{bradwellImplementingAffordableSocially2022}:

\begin{quote}
    \textit{We did have a lady that enjoyed feeding it. And she had a puree diet}, said a caregiver.
\end{quote}

Rouaix et al. \cite{rouaixAffectiveEngagementIssues2017} observed inquiries about what Nao eats, suggesting a projection of empathetic care onto the robots.
Thunberg et al. \cite{thunbergRobotPetsDecrease2020} also noted participants asking about emptying litter boxes, feeding it and letting it out, and described a situation where a person with dementia attempted to assist JfA Cat in sitting up, further demonstrating the tendency to perceive robots as requiring physical support.
Other caregiving behaviors include checking Hiro's diaper when it cried \cite{sumioka2021minimal}:  

\begin{quote}
\textit{``Participant C, who has severe dementia in the face group, repeatedly touched the robot's crotch and tried to undress it when it started crying. The staff member suggested that it looked like she was checking the robot's diaper''}, as reported by the author.  
\end{quote}  

Even when recognizing their limitations, PLwD demonstrate an awareness of the perceived needs of robots. For example, Sumioka et al. \cite{sumioka2021minimal} reported that the humanoid robot Hiro emitting a crying sound was perceived as an indication of hunger, prompting a person with dementia to respond with concern like toward a baby, saying,  

\begin{quote}
\textit{I can't breastfeed him.}
\end{quote}

\begin{quote}
\textit{``When the robot meowed or rolled over on its back, he thought that the robot was in pain or needed help. He asked it what was wrong and tried to help it sit up again,''} reported by the authors.  
\end{quote}

The awareness of their inability to respond to the robot's needs even caused distress. This was observed by Pike et al. \cite{pikeRobotCompanionCats2021}, who reported that some PLwD regarded meowing as the cat asking for something, and when they were unable to respond to the perceived need, it led to emotional distress.

\subsubsection{PLwD Perceive the Ontological Status of Social Robots in an Ambivalent and Fluctuating Way.}

The last theme is that the perception of social robots by PLwD is not static. Instead, it fluctuates between viewing the robot as a mechanical object and as a lifelike being.
Robotic behaviors, particularly motion, play a crucial role in shaping and sustaining the perception of lifelike qualities, and triggering this shift.  
As Bradwell et al. \cite{bradwellImplementingAffordableSocially2022} reported, although PLwD recognized the misalignment between the robot and real animals in terms of size and weight, behaviors such as turning and moving the head made JfA Dog appear more realistic.  
Gustafsson et al. \cite{gustafsson2015using} described a case in which a participant's perception of JustoCat shifted from day to day, recognizing it as robotic at times and at others as a real cat, influenced by its behaviors, especially purring and breathing.  
Similarly, caregivers in Thunberg et al. \cite{thunbergRobotPetsDecrease2020} noted that while most residents understood the robot was not a real animal, their perceptions often changed when it moved, leading them to believe it was real.  

This oscillation suggests a form of cognitive ambivalence, where truthful and untruthful perceptions coexist and shift dynamically. For example, Hung et al. \cite{hungExploringPerceptionsPeople2021} report a person with dementia who said to Paro ``\textit{oh my god, are you a friendly little seal? (singing cheerfully) I love you.}'' while partially recognizing its machine nature, ``\textit{I am scratching his tummy. There are his batteries. He's not dead yet.}''
Similarly, Pu et al. \cite{puHowPeopleDementia2020} report people with mild dementia chose to consider Paro as a real seal, while they were aware that it was a robot, saying ``\textit{It's like a real toy... Pretty much like a seal.}''

\section{Discussion}\label{sec:discussion}

\subsection{Summary of Results and Answers to RQ}\label{sec:discussion_summary}

This scoping review examined the design cues embedded in social robots used for PLwD and the perceptions and responses observed in PLwD during interactions with robots. 
First, the reviewed studies demonstrate that social robots for PLwD employ a wide range of verbal cues, including semantic speech and semantic-free speech, as well as nonverbal cues, including gestures, gaze, facial expression, and body and head movement (\textbf{RQ1a}). These cues are often combined in ways that resemble physiological signs (e.g. breathing, heartbeat, or blinking), social intentions (e.g. playful movement to invite interaction), and familiar beings (e.g., female appearance and voice, childlike gestures or cat-like movements). At the same time, certain design cues may also highlight the robot's artificial nature (e.g., mechanical movements or plush texture).

Second, the findings indicate that PLwD frequently perceive social robots as possessing biological traits, mental capacities, and attributes of social roles and categories, and show empathy and caretaking tendency towards them (\textbf{RQ2a}).
From the reviewed studies, we can also conclude that the ontological status of social robots is ambivalent and fluctuating in PLwD, oscillating between the perception of a robot as an artifact and a living being.

\begin{figure*}[h]%
\centering
\includegraphics[width=\linewidth]{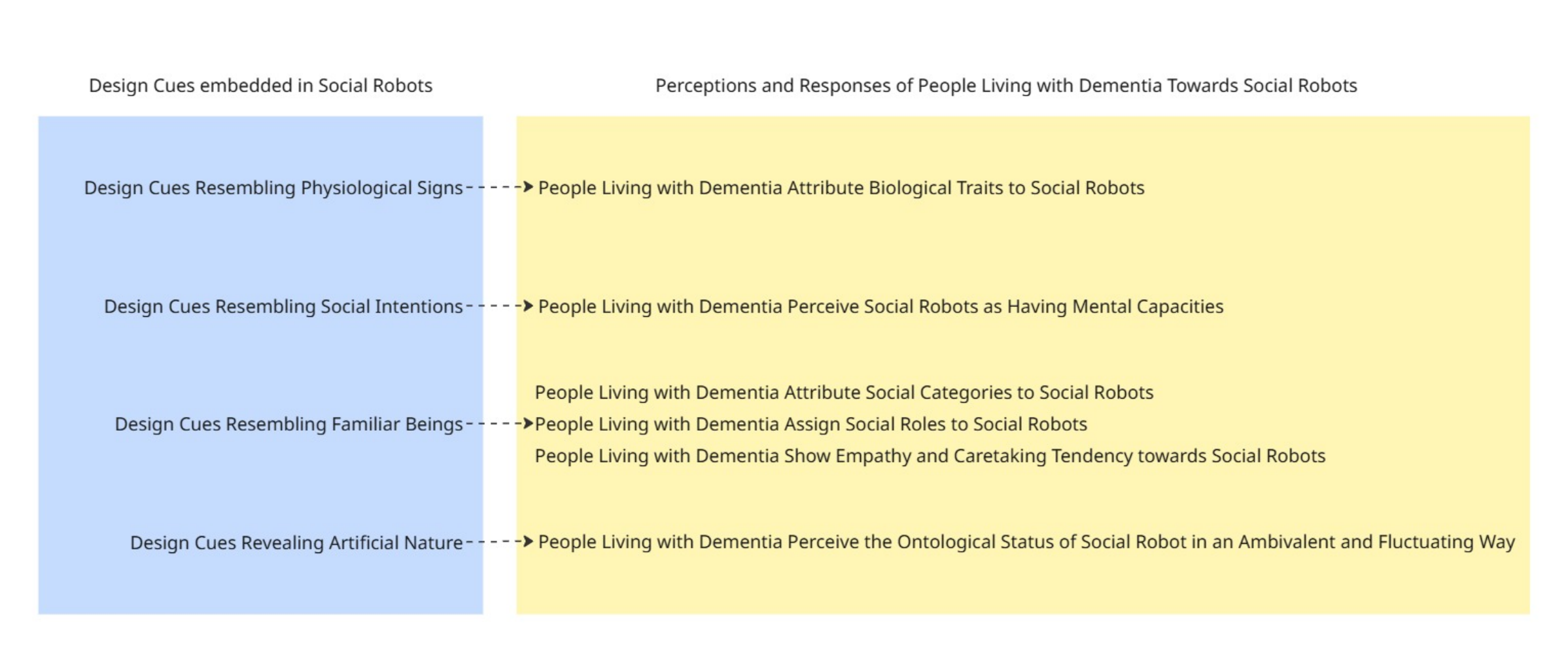}
\caption{Mind map illustrating reported design cue categories and documented perceptions and responses of PLwD in the reviewed studies}\label{themes}
\end{figure*}

We notice conceptual connections between the cues and the perceptions and responses we identified (See Figure~\ref{themes}), so there might be interdependencies between them worth examining in experimental studies. 
For example, cues resembling physiological signs (e.g., breathing, heartbeat, or blinking) may be interpreted as indicators of biological life (e.g., alive, sick, or dead). Cues that mimic social intentions (e.g., playful movement to invite interaction) could prompt users to attribute mental capacities, such as intention or agency.
Similarly, features that evoke familiar beings (e.g., female appearance and voice, childlike gestures, or seal-like head movements) can lead participants to assign social categories (e.g., gender or age), define social roles (e.g., pet or child), and even express empathy or caretaking behaviors (e.g., feeding or checking diaper).
Conversely, cues that expose the robot's artifactuality (e.g., mechanical movements or plush texture) can draw attention to its non-human status and reframe its ontological categorization.
The causal links between specific robot design cues and user perception cannot be formally established based on the evidence base in this literature review, so ours are mostly hypotheses.
Experimental studies, manipulating design cues and checking how these affect the perception of robots, might help establish such causal connections in a more capillary way.

\subsubsection{Design Cues of Social Robots (RQ1b)}
The reviewed studies suggest that false beliefs often arise from specific combinations of cues that add nuances of realism to each other. These cues are typically multimodal, combining embodiment, form, body structure, surface material, and sounds/speech. For example, non-speech vocalizations such as purring, behaviors such as the robotic cat rolling over its back, as well as specific appearances (e.g., soft cat fur) may add to one another and amplify PLwD's confabulation around the robots' lifelike nature.

A mechanism through which robot design cues seem to operate in this literature review is through enabling affordances. As Paauwe et al. \cite{paauwe2015designing} suggest, it is what social robots afford and their esthetic appearance that makes them perceived realistic and lovable in healthcare setting.
Many robots used for PLwD, such as JustoCat, JfA Cat/Dog, Paro, Pleo, Hiro, and Telenoid, are designed with sizes, materials, and shapes that afford PLwD's bodily interaction. They can be held, stroked, hugged, or carried in ways that resemble how people interact with pets and infants. These affordances may situate the robot within familiar relational scripts that involve companionship or care. From our results, such embodied interactions may also evoke autobiographical memories, reminding PLwD of the pets they adopted or their work experience \cite{feng2019livenature, joshi2019robots, pikeRobotCompanionCats2021, puHowPeopleDementia2020, yamazakiAcceptabilityTeleoperatedAndroid2014}. It might be specifically this reminiscence that transports the robots into the realm of lifelikeness, and makes PLwD more prone to interpret them as real living creatures, a hypothesis worth testing in future research.

Across the reviewed robots, another recurring design pattern is the extensive use of baby schema, kawaii design, and cuteness. Robots such as NAO, QT, Eva, Hiro, Paro, Pleo, JustoCat, Cuddler, Lovot, Miro, Furby and Sanne (see Figure~\ref{Robots}) share design characteristics including large eyes, rounded faces, oversized heads, simplified body proportions, soft textures, and baby- or pet-like vocalizations \cite{shiomi2025makes}.
Such cute design is known to elicit strong emotional responses and caregiving motivations \cite{perugia2026aww}, and has been described as a dark pattern when used to influence user behavior or limit conscious agency \cite{lacey2019cuteness}. In dementia care contexts, baby schema cues may orient PLwD toward interpreting robots as vulnerable or dependent beings, intensifying emotional engagement and perceived responsibility \cite{sumioka2021minimal, thunbergRobotPetsDecrease2020}. This could inadvertently contribute to SRD \cite{gray2018dark}. As cuteness responses are biologically rooted \cite{glocker2009baby}, the use of baby schema features in social robots for PLwD might constitute a dark pattern, as the conscious understanding of the robot's ontological nature, when available, might get overridden by the emotional responses the robot elicits.

As a last point, it is very interesting to notice that some design cues identified in this literature may signal the robot's artificial nature. We would like to propose that these machine-like characteristics, rather than representing failures of realism, could function as ethical design constraints, mitigating or countering PLwD's misleading interpretations of a robot's nature or capabilities and breaking the illusions of lifelikeness \cite{nielsen2026robot, lawrence2025role}.
As Henrik Skaug Sætra suggests, a key ethical aspect related to robot deception is to signal actual capabilities and to counteract human tendencies to anthropomorphize technology \cite{saetraSocialRobotDeception2021}.
While we call for future research developing ethically grounded design principles that balance transparency with SRD, we also suggest that one such principle could be this.

\subsubsection{Perceptions and Response of PLwD (RQ2b)}\label{sec:perception}

Attributing biological traits, mental capacities, social roles, and social categories to social robots, and showing empathy and caretaking tendency towards them are natural tendencies even in people without dementia. Such responses are common in HRI and do not necessarily provide evidence that users are misled about the nature and capabilities of the robot \cite{rosero2024human}. As Złotowski et al. \cite{zlotowski2015anthropomorphism} note, people may project deeply human attributes onto robots, ``\textit{even if these entities are nothing more than a plastic shell with an engine in it.}''

In many cases, users knowingly engage with social robots as if they were social beings while remaining aware that they are in fact artificial artifacts.
For example, Marchetti et al. noted that some residents chose whether to engage playfully with the robot as it is a cat or acknowledge its machine nature, retaining a sense of autonomy and dignity even while participating in a make-believe interaction \cite{marchetti2022pet}.
The phenomenon whereby users temporarily set aside the robot's artificiality in order to support emotional or relational engagement is often described as a ``willing suspension of disbelief'' \cite{saetraSocialRobotDeception2021, sharkeyWeNeedTalk2021}. In these situations, anthropomorphic responses are part of a voluntary interactional stance rather than evidence of deception.

Looking at our results and keeping our focus on PLwD's agency (as stated in our positionality), we argue that SRD becomes problematic when it does not stem from a willing suspension of disbelief on the part of PLwD but is instead unawared or unconscious. In these situations, the social contract stipulated between the person living with dementia and the robot has not been previously negotiated and agreed upon, and the individual with dementia thus engages in the interaction without awareness of its nature.

None of the existing definitions of SRD capture this aspect. Hence, we draw upon dual-process theory to sugges our own defintion. We describe deception as a dynamic, situated processing of social robots. According to dual-process theory \cite{frankish2010dual}, when a person interacts with the world, two distinct processing types are performed. Type 1 operates rapidly and unconsciously, relying on heuristic cues and pattern recognition to form intuitive judgments. In contrast, Type 2 operates more slowly and analytically, enabling deliberate reasoning, verification, and rationalization of initial impressions. In PLwD, the balance between Type 1 and Type 2 processing may fluctuate depending on the situation. PLwD may therefore come to rely more heavily on superficial social signals, such as the sound of a robot or its facial expressions, without critically evaluating the underlying mechanics or capabilities of the robot. As a result, they may intuitively form false beliefs about the robot's agency (e.g., the robot is a sentient being), intentions (e.g., the robot genuinely likes them), or roles (e.g., the robot is their child) and exhibit responses typically directed toward living beings (e.g., trying to breastfeed the robot and checking the diaper).
We thus argue that false beliefs in PLwD may arise when intuitive, heuristic-based interpretations of robot behavior (Type 1 processing) are not revised through more reflective reasoning (Type 2 processing). In all these situations, the suspension of disbelief is unconscious, and the SRD deriving from the interaction with the robot warrants a thorough examination of the clinical and therapeutic benefits it could bring about before taking place.
We argue that if therapeutic benefits greatly exceed the ethical risks, one could make the decision of using a social robot regardless of SRD. A critical question here, one to explore in future work, is who makes this decision and what interests might shape it \cite{wang2026following}.


\subsubsection{What Complicates the Picture}
As shown by this scoping review, various aspects may influence how PLwD perceive and respond to social robots, and thus complicate the detection of SRD. First, several studies indicate that people in the later stages of dementia may be more susceptible to deception than those in earlier stages. 
Furthermore, dementia itself is a fluctuating condition in which cognitive clarity oscillates. Even within the same individual, perceptions may vary due to disease progression and daily condition 
\cite{gustafsson2015using}.
In addition, broader contextual dynamics, such as how the robot is introduced (i.e., clear identification as robotic entity, neutral approach, or similarly to a living being) or framed (i.e., maintain the illusion of lifelikeness, redirect the PLwD's question about robot's nature, or explain robotic nature truthfully) by facilitators, may shape users' interpretations (See Section~\ref{sec:results_framingrobots}). 
In this sense, robotic deception is not merely a static feature of the robot itself, but a phenomenon that emerges through the evolving performances and narratives co-created between users, social robots, and other actors \cite{coeckelbergh2018describe}.

In dementia care, where the ontological status of the robot is often already ambiguous, these condition-related, relational and interactional aspects may further tilt interpretations by feeding Type 1 processing. 
We suggest that this can be mitigated in at least three ways highlighted by this literature review. First, by being aware of the baseline condition of the person with dementia on the day of the interaction. Second, by introducing the robot to PLWD following a hybrid of the neutral and mechanical approaches, hinting to the robot's functionalities (e.g., it has sensors in its whiskers) without falling into technicalities. Third, by following on deceptive perceptions from PLwD but avoiding to proactively feed them.

\subsection{A Methodological Note}\label{subsec5.4}
Our understanding of SRD in dementia care remains incomplete due to the lack of direct insights from PLwD. As Kant said, ``\textit{We do not have unmediated access to things as they are `in themselves' (noumena), but only to things as they appear to us (phenomena), shaped by the innate structures of our mind.}'' 
In the context of dementia care, this epistemological challenge is compounded by a practical observation: although many people with mild dementia retain the ability to articulate their own experiences, their perspectives are often overlooked \cite{suijkerbuijk2019active}.
In the reviewed papers, responses and perceptions of social robots are mostly filtered through the eyes of researchers, caregivers, or in general third-persons or proxies, and are thus not necessarily representative of PLwD's perspectives. Therefore, we highlight the necessity of validating the problematic nature of SRD, captured by our definition, with first person insights of PLwD. This could be done using narrative modes of qualitative inquiry.

\section{Conclusion}\label{sec:conclusion}




This scoping review examined how SRD may arise in interactions between social robots and PLwD by analyzing both robot design cues and user perceptions reported across 26 empirical studies. The findings show that PLwD frequently attribute biological traits, mental capacities, and social roles to robots. However, such interpretations do not necessarily indicate that users hold a literal false belief about the robot's nature. Building on these insights, we proposed a dual-process interpretation that distinguishes between everyday anthropomorphic engagement and situations in which users may genuinely misinterpret a robot's ontological status. This distinction helps clarify when SRD might occur and when lifelike interaction may instead support meaningful engagement. Future work should therefore examine how robot design, facilitation practices, and care contexts jointly shape how social robots are interpreted and experienced in dementia care.

\section*{Conflict of interest}
The authors have no conflicts of interest to declare that are relevant to the content of this article.

\section*{Ethics declaration}
Not applicable.

\section*{Authors' contributions} 
FW, GP, YF, and WI conceptualized this literature review and contributed to the development of the research questions and the framework. FW, GP, and YF contributed to the literature screening and analysis. FW prepared the original draft. GP and YF contributed to the writing, reviewing and editing. WI contributed to reviewing the final manuscript. GP supervised the whole process. All authors read and approved the final version.

\section*{Data Availability}
This article does not contain any original data. All data discussed are available in the publications cited in the reference list.

\section*{Statement of Using Generative AI}
Generative AI (ChatGPT 4o) was used for proofreading and polishing language.

\section*{Acknowledgments} 
The authors would like to acknowledge all those who contributed to this study for their time and support. We thank Claudia Sprenger for her assistance with data collection, Teis Arets for his help with the thematic analysis, Jiaxin Xu and Baisong Liu for their contributions to improving clarity and overall quality. 
This study is partially sponsored by Expertise Center for Dementia \& Technology (ECDT). Fan's work is funded by China Scholarship Council (202306010082). Giulia Perugia's work is partly funded by the research programme Ethics of Socially Disruptive Technologies (ESDiT), which is funded through the Gravitation programme of the Dutch Ministry of Education, Culture, and Science and the Netherlands Organization for Scientific Research (NWO grant number 024.004.031). Yuan's research is supported by the Humanity and Social Science Youth Foundation of Ministry of Education of China (25YJCZH048). Wijnand IJsselsteijn's work is supported through research project Quality of Life by use of Enabling AI in Dementia (QoLEAD), funded by NWO and Alzheimer Nederland.

\bibliographystyle{spmpsci}      
\bibliography{References.bib}   
\balance

\end{document}